\documentclass[
amsmath,
amssymb,
amsfonts,
aps,
nofootinbib,
preprintnumbers,
prl,
reprint,
superscriptaddress
]{revtex4-2}

\usepackage{xcolor}
\definecolor{MyDarkBlue}{rgb}{0.15,0.25,0.45} 
\usepackage[
linktocpage=true,
hypertexnames=false,
colorlinks=true,
citecolor=MyDarkBlue,
linkcolor=MyDarkBlue,
urlcolor=MyDarkBlue,
pdfauthor={
    Leron Borsten,
    Branislav Jurco,
    Hyungrok Kim,
    Tommaso Macrelli,
    Christian Saemann,
    Martin Wolf},
pdftitle={BRST-Lagrangian Double Copy of Yang--Mills Theory},
pdfsubject={hep-th},
breaklinks=true
]{hyperref}

\usepackage{amsthm}
\usepackage{mathtools}
\usepackage{bbm}
\usepackage{bm}
\usepackage{mathrsfs}
\usepackage{tikz}
\usetikzlibrary{matrix,cd,arrows}
\usepackage{ifthen}
\newcommand{\makecommand}[3]{%
    \foreach \i in #3 {%
        \expandafter\xdef\csname #1\i\endcsname{\noexpand#2{\unexpanded\expandafter{\i}}}%
    }%
}
\newcommand{\latinalphabet}{A,a,B,b,C,c,d,D,E,e,F,f,G,g,H,h,I,i,J,j,K,k,L,l,M,m,N,n,O,o,P,p,Q,q,R,r,S,s,T,t,U,u,V,v,W,w,X,x,Y,y,Z,z}
\makecommand{I}{\mathbbm}{\latinalphabet}
\makecommand{bf}{\mathbf}{\latinalphabet}
\makecommand{bm}{\bm}{\latinalphabet}
\makecommand{ca}{\mathcal}{\latinalphabet}
\makecommand{fr}{\mathfrak}{\latinalphabet}
\makecommand{rm}{\mathrm}{\latinalphabet}
\makecommand{sf}{\mathsf}{\latinalphabet}
\makecommand{sf}{\mathsf}{{id}}
\makecommand{sc}{\mathscr}{\latinalphabet}

\newcommand{\dpar}{\partial}
\newcommand{\wave}{\mathop\square}
\newcommand{\forw}{\uparrow}
\newcommand{\backw}{\downarrow}
\newcommand{\dbl}[1]{\boldsymbol{#1}}
\newcommand{\parder}[2][]{%
    \ifthenelse{\equal{#1}{}}{%
        \frac{\partial}{\partial #2}%
    }{%
        \frac{\partial #1}{\partial #2}%
    }%
}
\newcommand{\delder}[2][]{%
    \ifthenelse{\equal{#1}{}}{%
        \frac{\delta}{\delta #2}%
    }{%
        \frac{\delta #1}{\delta #2}%
    }%
}

\newtheoremstyle{breaknodot}
{\topsep}{\topsep}%
{\itshape}{}%
{\bfseries}{}%
{0pt}{\thmname{#1}\thmnumber{ #2.}~\thmnote{ \normalfont(#3)}}%
\theoremstyle{breaknodot}
\newtheorem{observation}{Observation}

\begin{document}
    
    \preprint{DMUS--MP--20/06}
    \preprint{EMPG--20--13}
    \title{BRST-Lagrangian Double Copy of Yang--Mills Theory}
    \author{Leron Borsten}
    \email[]{l.borsten@hw.ac.uk}
    \affiliation{Maxwell Institute for Mathematical Sciences\\ Department of Mathematics, Heriot--Watt University\\ Edinburgh EH14 4AS, United Kingdom}
    \author{Branislav Jur{\v c}o}
    \email[]{branislav.jurco@gmail.com}
    \affiliation{Charles University Prague\\ Faculty of Mathematics and Physics, Mathematical Institute\\ Prague 186 75, Czech Republic}
    \author{Hyungrok Kim}
    \email[]{hk55@hw.ac.uk}
    \affiliation{Maxwell Institute for Mathematical Sciences\\ Department of Mathematics, Heriot--Watt University\\ Edinburgh EH14 4AS, United Kingdom}
    \author{Tommaso Macrelli}
    \email[]{t.macrelli@surrey.ac.uk}
    \affiliation{Department of Mathematics, University of Surrey\\ Guildford GU2 7XH, United Kingdom}
    \author{Christian Saemann}
    \email[]{c.saemann@hw.ac.uk}
    \affiliation{Maxwell Institute for Mathematical Sciences\\ Department of Mathematics, Heriot--Watt University\\ Edinburgh EH14 4AS, United Kingdom}
    \author{Martin Wolf}
    \email[]{m.wolf@surrey.ac.uk}
    \affiliation{Department of Mathematics, University of Surrey\\ Guildford GU2 7XH, United Kingdom}
    
    \date{\today}
    
    \begin{abstract}
        We show that the double copy of gauge theory amplitudes to $\caN=0$ supergravity amplitudes extends from tree level to loop level. We first explain that color--kinematics duality is a condition for the Becchi--Rouet--Stora--Tyutin operator and the action of a field theory with cubic interaction terms to double copy to a consistent gauge theory. We then apply this argument to Yang--Mills theory, where color--kinematics duality is known to be satisfied onshell at the tree level. Finally, we show that the latter restriction can only lead to terms that can be absorbed in a sequence of field redefinitions, rendering the double copied action equivalent to $\caN=0$ supergravity.    
    \end{abstract}
    
    \maketitle
    
    \section{Introduction and summary}
    
    Yang--Mills scattering amplitudes have been conjectured to satisfy a color--kinematics (CK) duality~\cite{Bern:2008qj, Bern:2010ue, Bern:2010yg}: each amplitude can be written as a sum over purely trivalent graphs  such that the  kinematical numerators satisfy the  same antisymmetry and Jacobi identities as the color contributions. CK duality has been shown to hold at the tree level~\cite{Stieberger:2009hq,BjerrumBohr:2009rd,Jia:2010nz, BjerrumBohr:2010hn,Feng:2010my,  Chen:2011jxa, Mafra:2011kj, Du:2016tbc, Mizera:2019blq}. If it holds,  replacing  the color contributions of a Yang--Mills amplitude with another copy of the kinematical contributions yields a  gravity amplitude~\cite{Bern:2010yg}. This is known as the double copy prescription, and it has far reaching consequences for our understanding of quantum gravity. For reviews and  references see~\cite{Carrasco:2015iwa, Bern:2019prr,Borsten:2020bgv}. 
    
    Explicit nontrivial examples \cite{Bern:2010ue,Bern:2010tq,Carrasco:2011mn,Bern:2011rj,BoucherVeronneau:2011qv,Bern:2012cd,Bern:2012gh, Oxburgh:2012zr,Bern:2012uf,Du:2012mt,Yuan:2012rg,Bern:2013uka, Boels:2013bi,Bern:2013yya,Bern:2013qca, Bern:2014lha, Bern:2014sna, Mafra:2014gja, Mafra:2015mja, Johansson:2017bfl,  Bern:2017ucb, Bern:2018jmv}  have suggested that the double copy extends to the loop level (i.e.~to the integrands of loop amplitudes). In this letter, we argue that this is indeed the case to any finite loop order.
    
    Our approach builds on the ideas of manifestly CK-dual classical kinematic structure constants and Lagrangians~\cite{Bern:2010yg, Monteiro:2011pc, BjerrumBohr:2012mg,  Tolotti:2013caa,  Monteiro:2013rya, Fu:2016plh,  Cheung:2016prv, Chen:2019ywi, Borsten:2020xbt}. A key ingredient in our argument is the Becchi--Rouet--Stora--Tyutin (BRST) formalism and its enlarged field space of external states~\cite{Kugo:1977yx}. We extend the idea that the BRST framework can be double copied~\cite{Anastasiou:2014qba, Borsten:2017jpt, Anastasiou:2018rdx, Zoccali:2018thesis,Borsten:2019prq, Luna:2020adi, Borsten:2020xbt} and double copy the complete BRST Lagrangian. See also~\cite{Mafra:2014gja} for a powerful approach to loop-level CK-dual amplitudes using the BRST-invariance of the underlying pure spinor superstring.
    
    We make the crucial observation that onshell, CK duality violations due to longitudinal gluon modes can be compensated by harmless field redefinitions of the Nakanishi--Lautrup (NL) field. The Ward identities of the BRST symmetry then allow us to transfer CK duality from gluon amplitudes to those involving ghosts. Finally, onshell tree-level CK duality on the BRST-extended field space turns out to suffice to show that the BRST-Lagrangian double copied theory provides the loop integrands of a consistent perturbative quantization of $\caN=0$ supergravity. We stress that our results do not imply or rely on loop-level CK duality.
    
    A longer paper giving explicit expressions for many of the steps discussed only abstractly in the following and explaining the origin of the double copy in terms of homotopy algebras, mathematical objects unifying scattering amplitudes and BRST Lagrangians, is in preparation~\cite{Borsten:2021hua}. There, we also intend to make a connection to the observed non-trivial modifications of CK duality at the loop level, cf.~e.g.~\cite{Bern:2017yxu,Casali:2020knc}.

    \section{The BRST-Lagrangian double copy}\label{sec:BRST_Lag_double_copy}
    
    We start with an abstract perspective on the double copy. Any Lagrangian field theory is equivalent to a field theory with exclusively cubic interaction terms, by blowing up higher order vertices using auxiliary fields, cf.~also~\cite{Jurco:2018sby,Macrelli:2019afx}. A generic cubic action is
    \begin{equation}\label{eq:cubic_Lagrangian_standard_form}
        S=\frac{1}{2}\Phi^I\sfg_{IJ}\Phi^J+\frac{1}{3!}\Phi^I\sff_{IJK}\Phi^J\Phi^K,
    \end{equation}
    where the fields $\Phi^I$ are elements of some field space $\mathfrak{F}$ and the DeWitt index $I$ encodes all field labels (including position $x$). Summation and space-time integration over repeated indices are understood. We are interested in theories invariant under a gauge symmetry described by a BRST operator $Q$. 
    
    It is not hard to see that by blowing up ghost vertices in the Batalin--Vilkovisky (BV) action before gauge fixing, one can always reduce the gauge transformations of all fields to be at most cubic in the fields:
    \begin{equation}\label{eq:cubic_BRST_standard_form}
        Q\Phi^I=\sfq^I_J\Phi^J+\frac{1}{2}\sfq^I_{JK}\Phi^J\Phi^K+\frac{1}{3!}\sfq^I_{JKL}\Phi^J\Phi^K\Phi^L.
    \end{equation}
    We further require that fields split into ``left'' and ``right'' components (with independent left and right ghost numbers), but over a common space-time point. Consequently, we expand the DeWitt indices as $I=(\alpha,\bar\alpha,x)$, $J=(\beta,\bar\beta,y)$, and $K=(\gamma,\bar\gamma,z)$ and assume locality, so that we obtain
    \begin{subequations}
        \begin{equation}\label{eq:cubic_Lagrangian_coefficients_g}
            \sfg_{IJ}=\delta(x-y)\sfg_{\alpha\beta}(x)\bar{\sfg}_{\bar\alpha\bar\beta}(x)\wave ,
        \end{equation}
        \begin{equation}
            \begin{aligned}
                \sff_{IJK}\Phi^J\Phi^K&=\delta(x-y)\delta(x-z)\times
                \\
                &\hspace{0.5cm}\times\sum_{A,\bar A}(\sff^{A}_{\alpha\beta\gamma}\bar\sff^{\bar A}_{\bar\alpha\bar\beta\bar\gamma}\Phi^{\beta\bar \beta})
                (\sff^{'A}_{\alpha\beta\gamma}\bar\sff^{'\bar A}_{\bar\alpha\bar\beta\bar\gamma}\Phi^{\gamma \bar \gamma})
            \end{aligned}
        \end{equation}
        with $\sfg_{\alpha\beta}$ and $\bar{\sfg}_{\bar\alpha\bar\beta}$ graded (with respect to the ghost numbers) symmetric, and $\sff^{\delta A}_{\beta\gamma}$, etc., differential operators with constant coefficients. The indices $A$ and $\bar A$ range over the summands in $\sff_{IJK}$. To simplify notation, we define
        \begin{equation}\label{eq:cubic_Lagrangian_coefficients_f}
            \sff_{IJK}\Phi^J\Phi^K=:\sfg_{\alpha\delta}\bar\sfg_{\bar\alpha\bar\delta}\sff^\delta_{\beta\gamma}\bar\sff^{\bar\delta}_{\bar\beta\bar\gamma}\Phi^{\beta \bar \beta}\Phi^{\gamma\bar \gamma}.
        \end{equation}
    \end{subequations}
    Suppressing the position dependence, the Lagrangian of the theory becomes
    \begin{equation}\label{eq:cubic_Lagrangian_left_right}
        \caL=\frac12 \Phi^{\alpha\bar\alpha}\sfg_{\alpha\beta}\bar\sfg_{\bar\alpha\bar\beta}\wave\Phi^{\beta\bar\beta}+\frac1{3!}\Phi^{\alpha\bar\alpha}\sff_{\alpha\beta\gamma}\bar \sff_{\bar \alpha\bar \beta\bar \gamma} \Phi^{\beta\bar\beta}\Phi^{\gamma\bar\gamma},
    \end{equation}
    where we used the shorthand $\sff_{\alpha\beta\gamma}\sff_{\bar\alpha\bar\beta\bar\gamma}$ for the evident expression in~\eqref{eq:cubic_Lagrangian_coefficients_f}. 
    
    Analogously, we want the BRST operator to act on left and right indices separately, and we split $Q=Q_L+Q_R$ with 
    \begin{equation}\label{eq:cubic_BRST_left}
        \begin{aligned}
            Q_L\Phi^{\alpha\bar\alpha}=&\sfq^\alpha_\mu\delta^{\bar\alpha}_{\bar\mu}\Phi^{\mu\bar \mu}+\frac12\sfq^{\alpha}_{\mu\nu}\bar\sff^{\bar\alpha}_{\bar\mu\bar\nu}\Phi^{\mu\bar\mu} \Phi^{\nu\bar\nu}
            \\&
            +\frac1{3!} \sfq^{\alpha}_{\mu\nu\kappa}\bar \sff^\alpha_{\bar\mu\bar\nu\bar\kappa}\Phi^{\mu\bar\mu} \Phi^{\nu\bar\nu} \Phi^{\kappa\bar\kappa},
        \end{aligned}
    \end{equation}
    where $\bar\sff^{\bar\alpha}_{\bar\beta\bar\gamma\bar\delta}=3\bar\sff^{\bar\alpha}_{\bar\varepsilon\bar\delta}\bar\sff^{\bar\varepsilon}_{\bar\beta\bar\gamma}$  and similarly for $Q_R\Phi$. 
    
    As an example, consider the special case of cubic Yang--Mills theory, where the $\sfg_{\alpha\beta}$ and $\sff^\alpha_{\beta\gamma}$ are the components of the Killing form and the structure constants of a gauge algebra, respectively, while $\bar \sfg_{\bar\alpha\bar\beta}$ and $\bar \sff^{\bar \alpha}_{\bar \beta\bar \gamma}$ are the inner product and kinematical structure constants on the full BRST field space.

    To \emph{double copy} means to replace the left  (or right) sector with a copy of the right (or left) sector of some, not necessarily the same, theory written in the form~\eqref{eq:cubic_Lagrangian_left_right},~\eqref{eq:cubic_BRST_left}. If the resulting action $\dbl{S}$ and BRST operator $\dbl{Q}$ satisfy again the relations $\dbl{Q}^2=0,$ $\dbl{Q}\dbl{S}=0$, we obtain a consistently gauge-fixed theory ready for quantization.
    
    It is not hard to see that $\dbl{Q}_{L/R}^2=0$ iff $Q_{L/R}^2=0$; the condition $\dbl{Q}_L\dbl{Q}_R+\dbl{Q}_R\dbl{Q}_L=0$ may induce further conditions. For Yang--Mills theory, one readily computes that CK duality suffices for the condition $\dbl{Q}\dbl{S}=0$. If CK duality fails to hold up to certain terms, then $\dbl{Q}\dbl{S}=0$ also fails to hold up to the same terms, possibly multiplied by other fields and their derivatives. (Mathematically, the terms describing the failure of CK duality generate an ideal in the algebra of fields and their derivatives. The expressions $\dbl{Q}\dbl{S}$ and $\dbl{Q}^2$ take values in this ideal.)

    \section{Preliminary observations}
    
    We are interested in perturbative aspects and omit any non-perturbative issues. Also, we are interested in $n$-point amplitudes up to $\ell$ loops for $n$ and $\ell$ finite. Thus, there is always a number $N\in\IN$ so that monomials of degree $m>N$ can be neglected in the Lagrangian. We always use the term ``amplitude'' for onshell states and the term ``correlator'' for offshell states.       
    
    Although the quantization of Yang--Mills theory does not require it, it is convenient to start from the BV form~\cite{Batalin:1981jr} of the Yang--Mills Lagrangian on Minkowski space, using canonical notation for all fields,
    \begin{equation}\label{eq:BVYMAction}
        \caL_{\rm YM}\coloneqq-\frac14 F^a_{\mu\nu}F^{a\mu\nu}+A^{+a}_\mu(\nabla^\mu c)^a+\frac{g}2f^a_{bc}c^{+a}c^bc^c+b^a\bar c^{+a},
    \end{equation}
    with $g$ the Yang--Mills coupling constant. We use the gauge fixing fermion $\Psi\coloneqq\Psi_0+\Psi_1$ with 
    \begin{equation}\label{eq:GFF}
        \Psi_0\coloneqq\int\rmd^dx\,\bar c^a\left(\frac{\xi}2b^a-\partial^\mu A^a_\mu\right),~~
        \Psi_1\coloneqq\int\rmd^dx\,\bar c^a\psi^a,
    \end{equation}
    where $\psi^a$ is of ghost number $0$ and depends at least quadratically on the fields and their derivatives. We obtain the gauge-fixed Lagrangian
    \begin{equation}\label{eq:YM_Lagrangian_0}
        \begin{aligned}
            \caL_{\rm YM}^{\rm gf} &= -\frac14 F^a_{\mu\nu}F^{a\mu\nu}-\bar c^a\partial^\mu(\nabla_\mu c)^a+\frac{\xi}2(b^a)^2-b^a \partial^\mu A^a_\mu
            \\
            &\kern1cm+\delder[\Psi_1]{A^a_\mu}(\nabla_\mu c)^a+\frac{g}2f^a_{bc}\delder[\Psi_1]{c^a}c^b c^c+b^a\delder[\Psi_1]{\bar c^a}.
        \end{aligned}
    \end{equation}
    For $\psi^a=0$, we recover the $R_\xi$-gauges. The BRST transformations are
    \begin{equation}\label{eq:BVYMOperator}
        \begin{aligned}
            Q_{\rm YM}c^a &\coloneqq -\frac{g}{2}f^a_{bc}c^bc^c,~~
            &Q_{\rm YM}A^a_\mu &\coloneqq (\nabla_\mu c)^a,
            \\
            Q_{\rm YM}b^a &\coloneqq 0,~~
            &Q\bar c^a &\coloneqq b^a,
        \end{aligned}
    \end{equation}
    satisfying $Q_{\rm YM}^2=0$ offshell. 
    
    The non-physical fields enlarge the one-particle field space of asymptotic onshell states by four types of states: the two unphysical polarizations of the gluon, called {\em forward} and {\em backward} and denoted by $A^\forw$ and $A^\backw$, and the ghost and antighost states~\cite{Kugo:1977yx}. All amplitudes will be built from the $n$-particle form of this BRST-extended onshell field space, which carries an action of the linearization of~\eqref{eq:BVYMOperator} denoted by $Q^{\rm lin}_{\rm YM}$. The physical polarizations are singlets, $Q^{\rm lin}_{\rm YM} A_{\rm phys}=0$, and we have two more doublets:
    \begin{equation}
        A^\forw~\xrightarrow{~Q^{\rm lin}_{\rm YM}~} c~~~\mbox{and}~~~\bar c \xrightarrow{~Q^{\rm lin}_{\rm YM}~} b=\frac{1}{\xi}\dpar^\mu A^\backw_\mu{+\dotsb},
    \end{equation}
    where the ellipsis indicates terms that arise from $\Psi_1$.
    
    \begin{observation}\label{ob:BRST_connected}
        The set of connected correlation functions is BRST-invariant because they can be written as linear combinations of products of correlation functions.
    \end{observation}
    
    Crucial to our discussion are the supersymmetric Ward identities generated by the BRST operator. We start with the onshell form, see~e.g.~\cite{Elvang:2013cua,Elvang:2015rqa}. Since the free vacuum is invariant under the action of $Q^{\rm lin}_{\rm YM}$, we have the following onshell Ward identities:
    \begin{equation}\label{eq:on_shell_Ward}
        0=\langle 0| [Q^{\rm lin}_{\rm YM},{\caO_1\dotsm \caO_n}]|0\rangle.
    \end{equation}
    
    We now consider the onshell Ward identity for ${\caO_1\dotsm \caO_n}= A^\forw\bar c (c\bar c)^k A^{n-2k-2}_{\rm phys}$ and obtain
    \begin{equation}
        \langle 0|(c\bar c)^{k+1}A_{\rm phys}^{n-2k-2}|0\rangle\sim\langle 0|A^\forw (c\bar c)^k b A_{\rm phys}^{n-2k-2}|0\rangle.
    \end{equation}
    Thus:
    \begin{observation}\label{ob:ghost_pair_reduction}
        Any amplitude with $k+1$ ghost--antighost pairs and all gluons transversely polarized is given by a sum of amplitudes with $k$ ghost pairs.
    \end{observation}
    From the construction of amplitudes via Feynman diagrams, it follows that we also have the following onshell Ward identity for an approximate BRST symmetry.
    \begin{observation}\label{ob:approximate_Ward}
        Suppose that $QS=0$ and $Q^2=0$ only onshell. Then, we still have~\eqref{eq:on_shell_Ward} together with a corresponding identification of amplitudes with $k+1$ ghost--antighost pairs and all gluons transversely polarized and a sum of amplitudes with $k$ ghost pairs.
    \end{observation}
    
    We shall also need the offshell form of these Ward identities, 
    \begin{equation}\label{eq:off_shell_Ward}
        \begin{aligned}
            \partial^\mu\langle &j_\mu(x) \caO_1(x_1){\dotsm} \caO_n(x_n)\rangle=\\
            &\sum_{i=1}^n\mp\delta (x-x_i)\langle (Q\caO_i(x_i))\Pi_{j\neq i}\caO_j(x_j)\rangle,
        \end{aligned}
    \end{equation}
    where $j_\mu$ is the BRST current. The left-hand side vanishes after integration over $x$, and using Observation~\ref{ob:BRST_connected}, we can restrict to connected correlators at a particular order in the coupling constant $g$ and then further to lowest order in $\hbar$, i.e.~to tree level. Consider now operators $\caO_i(x_i)$ for those  restricted Ward identities which are linear in the fields.
    \begin{observation}\label{ob:offshell-Ward}
        The onshell relations between tree amplitudes from Observation~\ref{ob:ghost_pair_reduction}~induced by~\eqref{eq:on_shell_Ward} extend to (offshell) tree-level connected correlators. For example, 
        \begin{equation}
            \begin{aligned}
                &\langle A_\mu(x_1) b(x_2) A_\nu(x_3)\rangle=\\
                &\hspace{0.5cm}\langle \dpar_\mu c(x_1) \bar c(x_2) A_\nu(x_3)\rangle+\langle A_\mu(x_1) \bar c(x_2) \dpar_\nu c(x_3)\rangle.
            \end{aligned}            
        \end{equation}
    \end{observation}
    
    Next, we make the following three general observations:
    \begin{observation}\label{ob:classical_equivalence}
        If two field theories have the same tree amplitudes, then the minimal models of their $L_\infty$-algebras coincide, cf.~\cite{Jurco:2018sby,Macrelli:2019afx}. If they have the same field content and kinetic parts, then they are related by a local (invertible) field redefinition. 
    \end{observation}
    
    \begin{observation}\label{ob:quantum_equivalence}
        Two field theories are quantum equivalent, if all their correlators agree. Since correlators can be glued together from tree-level correlators (up to regularization issues), it suffices if the latter agree. 
    \end{observation}
    
    \begin{observation}\label{ob:field_shifts}
        A shift of a field by products of fields and their derivatives which do not involve the field itself does not change the path integral measure. Local field redefinitions {that are trivial at linear order} produce a Jacobian that is regulated to unity in dimensional regularization~\cite{Tyutin:2000ht,tHooft:1973wag,Leibbrandt:1975dj}, see also~\cite{Henneaux:1992}. Therefore, they preserve quantum equivalence.
    \end{observation}
    
    In our constructions, we will also exploit the possibility of adjusting our choice of gauge. Performing shifts $b^a\mapsto b^a+X^a$ and $\Psi_1\mapsto\Psi_1+\Xi_1$ with $\Xi_1\coloneqq\int\rmd^dx\,\bar c^aY^a$ induces a shift of~\eqref{eq:YM_Lagrangian_0} by
    \begin{equation}\label{eq:add_terms}
        \begin{aligned}
            &\frac{\xi}2(X^a)^2+X^a(\xi b^a-\dpar^\mu A_\mu^a)+X^a\delder[\Psi_1]{\bar c^a}
            \\
            &+\delder[\Xi_1]{A^a_\mu}(\nabla_\mu c)^a+\frac g2f^a_{bc}\delder[\Xi_1]{c^a}c^b c^c+(b^a+X^a)\delder[\Xi_1]{\bar c^a}.
        \end{aligned}
    \end{equation}
    If $X^a$ is independent of the NL field $b^a$, this modification preserves the theory at the quantum level by Observation~\ref{ob:field_shifts}. Furthermore, if $X^a$ is at least quadratic in the fields, this transformation preserves the action of $Q^{\rm lin}_{\rm YM}$ on the BRST-extended onshell field space. 
    
    Consider now the special case $\psi^a=0$ and $X^a$ independent of $b^a$ and fix $Y^a$ iteratively such that the terms linear in $b^a$ of~\eqref{eq:add_terms} vanish:
    \begin{equation}
        \xi X^a+\delder[\Xi_1]{\bar c^a}=\xi X^a+Y^a+\bar c^b \parder[Y^a]{\bar c^b}{+\dotsb}=0~.
    \end{equation}
    This leads to the following observation:
    \begin{observation}\label{obs:b-field-shifts}
        Terms in the Lagrangian of the form $(\dpar^\mu A_\mu)^aX^a$ with $X^a$ at least quadratic in the fields and their derivatives but independent of the NL field can be removed in $R_\xi$-gauges by shifting the NL field. This creates additional terms~\eqref{eq:add_terms} which are at least of fourth order and preserve the amplitudes by Observation~\ref{ob:field_shifts}.
    \end{observation}
    \begin{observation}\label{ob:match_Nakanishi}
        Terms in the action that are proportional to a NL field can be absorbed by choosing a suitable term $\psi^a$. This leaves the physical sector invariant but it may modify the ghost sector. Because NL fields appear as trivial pairs in the BV action, it is not hard to see that this extends to general gauge theories, e.g.~with several NL fields and ghosts-for-ghosts.
    \end{observation}
    
    We also make the following three observations regarding the double copy.
    
    \begin{observation}\label{ob:usual_double_copy}
        The tree amplitudes of Yang--Mills theory can be written in CK-dual form~\cite{Stieberger:2009hq, BjerrumBohr:2009rd, Jia:2010nz, BjerrumBohr:2010hn, Feng:2010my, Chen:2011jxa, Mafra:2011kj, Du:2016tbc, Mizera:2019blq}.  
    \end{observation}
    
    \begin{observation}\label{ob:ampltiudes_to_Lagrangian}
        For amplitudes in CK-dual form, there is a corresponding local, cubic, and physically equivalent  Lagrangian whose partial amplitudes produce the kinematical numerators~\cite{Tolotti:2013caa}. 
    \end{observation}
    
    \begin{observation}\label{ob:established_dc}
        Double copying the Yang--Mills tree amplitudes in CK-dual form yields the tree amplitudes of $\caN=0$ supergravity~\cite{Bern:2008qj, Bern:2010ue,Bern:2010yg}.
    \end{observation}

    \section{CK-dual Yang--Mills theory}
    
    In order to BRST-Lagrangian double copy Yang--Mills theory, we first must bring its action into the normalized form~\eqref{eq:cubic_Lagrangian_left_right}. Our goal will be to construct abstractly a Lagrangian that guarantees tree-level CK duality for the BRST-extended onshell field space.
    
    CK duality of the Feynman diagrams for the field space of physical gluons can be guaranteed by adding terms to the Lagrangian~\cite{Bern:2010yg,Tolotti:2013caa} and subsequently {\em strictifying} these, i.e.~introducing a set of auxiliary fields such that all interaction vertices are cubic. This strictification is mostly determined by the color and momentum structure of the additional terms in the Lagrangian.
    
    It remains to ensure CK duality for tree amplitudes involving ghosts or backward polarized gluon states, which we do by introducing compensating terms, preserving quantum equivalence. (Forward polarized gluons can be absorbed by residual gauge transformations and therefore do not appear in the Lagrangian. Thus, they cannot contribute to CK duality violations.) 
    
    We implement the necessary changes iteratively for $n$-point amplitudes, starting with $n=4$. We can compensate for CK duality violations due to backward polarized gluons, which can be done by introducing terms of the form $(\dpar^\mu A_\mu)^aX^a$. By Observation~\ref{obs:b-field-shifts}, we can produce such terms, preserving quantum equivalence, and we immediately compensate for the additional terms linear in the NL field using Observation~\ref{ob:match_Nakanishi}. Since we perform all shifts at the level of the BV action and the gauge fixing fermion, the resulting action is automatically BRST-invariant and its amplitudes are CK-dual for external legs of ghost number $0$. 
    
    By Observation~\ref{ob:ghost_pair_reduction}, these amplitudes fully determine all amplitudes with ghosts and antighosts on external legs. Moreover, the CK-dual form of the former can be copied over to the latter, by literally copying trivalent Feynman diagrams for the gluon modes linked by the BRST symmetry to the ghost--antighost pairs. We do this iteratively in the number of ghost--antighost pairs. The consistency of the copying process is guaranteed by the full BRST symmetry of the action. We then use Observation~\ref{ob:established_dc} to turn these CK-dual amplitudes for arbitrary ghost number into a local, cubic, and BRST-invariant Lagrangian.
    
    The resulting Lagrangian $\caL_{\rm YM}^{\rm CK}$ is of the form~\eqref{eq:cubic_Lagrangian_left_right} and quantum equivalent to the Lagrangian $\caL_{\rm YM}$ given in~\eqref{eq:YM_Lagrangian_0}.
    
    \section{The BRST-Lagrangian double copy of Yang--Mills Theory}
    
    We now turn to the $\caN=0$ supergravity side. The gauge-fixed BRST Lagrangian $\caL_{\caN=0}$ of this theory is readily constructed. The following two diagrams concisely summarize the theory's field content from the perspective of the double copy, describing the symmetrized and antisymmetrized tensor products of two copies of the BRST Yang--Mills fields:
    
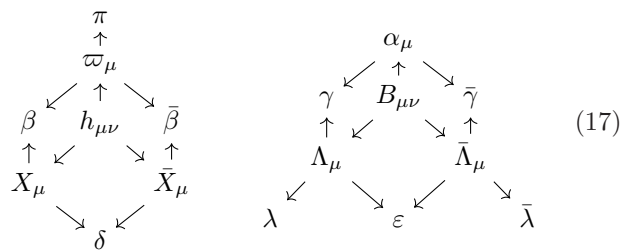
\begin{equation}\label{eq:field_content}
        \begin{tikzpicture}[baseline={(current bounding box.center)}]
            \matrix (m) [matrix of nodes, ampersand replacement=\&, column sep = 0.2cm, row sep = 0.2cm]{
                {} \& $\pi$ \& {} \\
                {} \& $\varpi_\mu$ \& {} \\
                $\beta$ \& $h_{\mu\nu}$ \& $\bar \beta$ \\
                $X_\mu$ \& {} \& $\bar X_\mu$ \\
                {} \& $\delta$ \& {} \\
            };
            \draw[->] (m-2-2) -- (m-1-2) ;
            \draw[->] (m-2-2) -- (m-3-1) ;
            \draw[->] (m-2-2) -- (m-3-3) ;
            \draw[->] (m-3-2) -- (m-2-2) ;
            \draw[->] (m-3-2) -- (m-4-1) ;
            \draw[->] (m-3-2) -- (m-4-3) ;
            \draw[->] (m-4-1) -- (m-3-1) ;
            \draw[->] (m-4-3) -- (m-3-3) ;
            \draw[->] (m-4-1) -- (m-5-2) ;
            \draw[->] (m-4-3) -- (m-5-2) ;
        \end{tikzpicture}
        \hspace{0.5cm}
        \begin{tikzpicture}[baseline={(current bounding box.center)}]
            \matrix (m) [matrix of nodes, ampersand replacement=\&, column sep = 0.2cm, row sep = 0.2cm]{
                {} \& {} \& $\alpha_\mu$ \& {} \& {}\\
                {} \& $\gamma$ \& $B_{\mu\nu}$ \& $\bar\gamma$ \& {}\\
                {} \& $\Lambda_\mu$ \& {} \& $\bar \Lambda_\mu$ \& {}\\
                $\lambda$ \& {} \& $\varepsilon$ \& {} \& $\bar \lambda$\\
            };
            \draw[->] (m-1-3) -- (m-2-2) ;
            \draw[->] (m-1-3) -- (m-2-4) ;
            \draw[->] (m-2-3) -- (m-1-3) ;
            \draw[->] (m-2-3) -- (m-3-2) ;
            \draw[->] (m-2-3) -- (m-3-4) ;
            \draw[->] (m-3-2) -- (m-2-2) ;
            \draw[->] (m-3-2) -- (m-4-1) ;
            \draw[->] (m-3-2) -- (m-4-3) ;
            \draw[->] (m-3-4) -- (m-2-4) ;
            \draw[->] (m-3-4) -- (m-4-3) ;
            \draw[->] (m-3-4) -- (m-4-5) ;
        \end{tikzpicture}
    \end{equation}
    Here, the physical fields of ghost number~0 are $h_{\mu\nu}$ (containing the metric perturbation about the Minkowski vacuum and the dilaton) and $B_{\mu\nu}$ (the Kalb--Ramond two-form). Ghost number increases by column from left to right, and all vector/form indices are made explicit. The arrows indicate factorization relations between the various fields~\cite{Borsten:2021hua}. In addition to the expected BRST field content, we have two trivial BV pairs $(\delta,\beta)$ and $(\bar \beta,\pi)$, see e.g.~\cite{Baulieu:2020obv} for the same fields in a different context. For more details, see~\cite{Borsten:2021hua} as well as~\cite{Anastasiou:2014qba, Borsten:2017jpt, Anastasiou:2018rdx, Zoccali:2018thesis, Borsten:2020xbt}.
    
    The double copy of $Q_{\rm YM}$ and $\caL^{\rm CK}_{\rm YM}$ yields a BRST operator $\dbl{Q}$ which satisfies $\dbl{Q}^2=0$ onshell and a Lagrangian $\dbl{\caL}$ for the field content~\eqref{eq:field_content}. The latter is quantum equivalent to the manifestly CK-dual, cubic or strict form $\caL^{\rm st}_{\caN=0}$ of $\caN=0$ supergravity obtained from Observation~\ref{ob:ampltiudes_to_Lagrangian}:
    
    \paragraph{(i) Kinematic equivalence:} The two kinematic Lagrangians are equivalent and linked by evident suitable field redefinitions~\cite{Borsten:2021hua}. The existence of such a field redefinition is ensured by the linear double copy BRST operator $\dbl{Q}^{\rm lin}$~\cite{Anastasiou:2018rdx, Borsten:2020xbt}, which is equivalent to the  linear BRST operator $Q_{\caN=0}^{\rm st,\,lin}$ and annihilates the quadratic double copy Lagrangian~\cite{Borsten:2021hua}. We implement the field redefinition on $\caL^{\rm st}_{\caN=0}$, obtaining $\caL^{\rm st,\,1}_{\caN=0}$.
    
    \paragraph{(ii) Ghost number 0, partly:}  Since the classical Yang--Mills action was written in a form with purely cubic, local interactions with manifest CK duality to all points, the tree amplitudes of $\dbl{\caL}$ for physical fields match those of $\caL^{\rm st,\,1}_{\caN=0}$, cf.~Observation~\ref{ob:established_dc}. The amplitudes for auxiliaries of ghost number $0$ are determined by collinear limits of amplitudes of physical fields and thus also agree between the theories. By Observation~\ref{ob:classical_equivalence}, we can implement a field redefinition $\caL^{\rm st,\,1}_{\caN=0}\rightarrow \caL^{\rm st,\,2}_{\caN=0}$ such that the interaction vertices of $\caL^{\rm st,\,2}_{\caN=0}$ and $\dbl{\caL}$ agree for physical and auxiliary fields of ghost number $0$ to any finite order. For these fields, also the tree-level correlators agree, because the field redefinitions preserve quantum equivalence by Observation~\ref{ob:field_shifts}.
    
    \paragraph{(iii) Gauge fixing sector:} The difference between $\dbl{\caL}$, after integrating out all auxiliary fields, and $\caL^{\rm st,\,2}_{\caN=0}$ proportional to any of the NL-like fields ($\beta,\bar \beta,\varpi_\mu,\pi,\gamma,\alpha_\mu,\bar \gamma$) can be absorbed in a choice of gauge fixing which will only create new terms in the ghost sector, cf.~Observation~\ref{ob:match_Nakanishi} for all fields except for $\beta$, which requires a slightly different treatment~\cite{Borsten:2021hua}. We implement this new gauge fixing, and take over the strictification from $\dbl{\caL}$, obtaining $\caL^{\rm st,\,3}_{\caN=0}$ together with a BRST operator $Q^{\rm st,\,3}_{\caN=0}$.
    
    \paragraph{(iv) Ghost sector:} Starting from the latter, we now use Observation~\ref{ob:ghost_pair_reduction} to copy over the CK-dual form of the amplitudes with external legs labeled exclusively by fields of ghost number $0$ to a CK-dual form of amplitudes with ghost--antighost pairs on external legs. This proceeds just as in the case of Yang--Mills theory and consistency is again guaranteed by full BRST symmetry of the action. By Observation~\ref{ob:established_dc}, we can then turn these CK-dual amplitudes into a local, cubic, and CK-duality manifesting Lagrangian $\caL^{\rm CK}_{\caN=0}$ physically equivalent to $\caL^{\rm st,\,3}_{\caN=0}$ and thus to $\caL_{\caN=0}$.
    
    Both $\dbl{\caL}$ and $\caL^{\rm CK}_{\caN=0}$ are local and have the same field content. The tree-level correlators involving physical and NL fields agree. Using the approximate Ward identities, cf.~Observation~\ref{ob:approximate_Ward}, and the fact that $\dbl{Q}^{\rm lin}$ and $Q^{\rm CK,lin}_{\caN=0}$ agree, we deduce that all tree amplitudes involving ghosts and antighost pairs agree, too. By construction, this agreement extends to individual onshell Feynman diagrams, between the strictifications $\dbl{\caL}$ and $\caL^{\rm CK}_{\caN=0}$, even for auxiliary fields: we can iteratively split off external vertices with two external legs, exposing Feynman diagrams with onshell external but offshell auxiliary fields. Up to a field redefinition of the auxiliaries, these also must agree. 
    
    The only potential remaining difference between $\dbl{\caL}$ and $\caL^{\rm CK}_{\caN=0}$ is then interaction terms containing $\wave \Gamma$ and $\wave \bar \Gamma$ terms for $\Gamma$ a ghost field. Going through the construction, one can argue that such terms, if they are there, have to appear in the same way in $\dbl{\caL}$ and $\caL^{\rm CK}_{\caN=0}$. Alternatively, one can show that both theories satisfy the same Ward identities for tree-level correlators, rendering them quantum equivalent by Observation~\ref{ob:quantum_equivalence}. The simplest argument, however, is to use Observation~\ref{ob:classical_equivalence} to note that both theories are related by a local field redefinition. Observation~\ref{ob:field_shifts} then implies that both theories are quantum equivalent.
    
    \

    \noindent\emph{Data Management.} No additional research data beyond the data presented and cited in this work are needed to validate the research findings in this work.
    
    \

    \begin{acknowledgments}
        \noindent \emph{Acknowledgments.}  We gratefully acknowledge stimulating conversations with Johannes Br{\"o}del, Michael Duff, Henrik Johansson, Silvia Nagy, Jim Stasheff, and Alessandro Torrielli. L.B., H.K., and C.S.~were supported by the Leverhulme Research Project Grant RPG-2018-329 ``The Mathematics of M5-Branes.'' B.J.~was supported by the GA\v{C}R Grant EXPRO 19-28268X and thanks MPIM Bonn for hospitality. T.M.~was partially supported by the EPSRC Grant EP/N509772. 
    \end{acknowledgments}
    

\begin{thebibliography}{64}%
\makeatletter
\providecommand \@ifxundefined [1]{%
 \@ifx{#1\undefined}
}%
\providecommand \@ifnum [1]{%
 \ifnum #1\expandafter \@firstoftwo
 \else \expandafter \@secondoftwo
 \fi
}%
\providecommand \@ifx [1]{%
 \ifx #1\expandafter \@firstoftwo
 \else \expandafter \@secondoftwo
 \fi
}%
\providecommand \natexlab [1]{#1}%
\providecommand \enquote  [1]{``#1''}%
\providecommand \bibnamefont  [1]{#1}%
\providecommand \bibfnamefont [1]{#1}%
\providecommand \citenamefont [1]{#1}%
\providecommand \href@noop [0]{\@secondoftwo}%
\providecommand \href [0]{\begingroup \@sanitize@url \@href}%
\providecommand \@href[1]{\@@startlink{#1}\@@href}%
\providecommand \@@href[1]{\endgroup#1\@@endlink}%
\providecommand \@sanitize@url [0]{\catcode `\\12\catcode `\$12\catcode
  `\&12\catcode `\#12\catcode `\^12\catcode `\_12\catcode `\%12\relax}%
\providecommand \@@startlink[1]{}%
\providecommand \@@endlink[0]{}%
\providecommand \url  [0]{\begingroup\@sanitize@url \@url }%
\providecommand \@url [1]{\endgroup\@href {#1}{\urlprefix }}%
\providecommand \urlprefix  [0]{URL }%
\providecommand \Eprint [0]{\href }%
\providecommand \doibase [0]{https://doi.org/}%
\providecommand \selectlanguage [0]{\@gobble}%
\providecommand \bibinfo  [0]{\@secondoftwo}%
\providecommand \bibfield  [0]{\@secondoftwo}%
\providecommand \translation [1]{[#1]}%
\providecommand \BibitemOpen [0]{}%
\providecommand \bibitemStop [0]{}%
\providecommand \bibitemNoStop [0]{.\EOS\space}%
\providecommand \EOS [0]{\spacefactor3000\relax}%
\providecommand \BibitemShut  [1]{\csname bibitem#1\endcsname}%
\let\auto@bib@innerbib\@empty
\bibitem [{\citenamefont {Bern}\ \emph {et~al.}(2008)\citenamefont {Bern},
  \citenamefont {Carrasco},\ and\ \citenamefont {Johansson}}]{Bern:2008qj}%
  \BibitemOpen
  \bibfield  {author} {\bibinfo {author} {\bibfnamefont {Z.}~\bibnamefont
  {Bern}}, \bibinfo {author} {\bibfnamefont {J.~J.~M.}\ \bibnamefont
  {Carrasco}},\ and\ \bibinfo {author} {\bibfnamefont {H.}~\bibnamefont
  {Johansson}},\ }\bibfield  {title} {\bibinfo {title} {New relations for
  gauge-theory amplitudes},\ }\href
  {https://doi.org/10.1103/PhysRevD.78.085011} {\bibfield  {journal} {\bibinfo
  {journal} {Phys. Rev. D}\ }\textbf {\bibinfo {volume} {78}},\ \bibinfo
  {pages} {085011} (\bibinfo {year} {2008})},\ \Eprint
  {https://arxiv.org/abs/0805.3993} {arXiv:0805.3993 [hep-ph]} \BibitemShut
  {NoStop}%
\bibitem [{\citenamefont {Bern}\ \emph
  {et~al.}(2010{\natexlab{a}})\citenamefont {Bern}, \citenamefont {Carrasco},\
  and\ \citenamefont {Johansson}}]{Bern:2010ue}%
  \BibitemOpen
  \bibfield  {author} {\bibinfo {author} {\bibfnamefont {Z.}~\bibnamefont
  {Bern}}, \bibinfo {author} {\bibfnamefont {J.~J.~M.}\ \bibnamefont
  {Carrasco}},\ and\ \bibinfo {author} {\bibfnamefont {H.}~\bibnamefont
  {Johansson}},\ }\bibfield  {title} {\bibinfo {title} {Perturbative quantum
  gravity as a double copy of gauge theory},\ }\href
  {https://doi.org/10.1103/PhysRevLett.105.061602} {\bibfield  {journal}
  {\bibinfo  {journal} {Phys. Rev. Lett.}\ }\textbf {\bibinfo {volume} {105}},\
  \bibinfo {pages} {061602} (\bibinfo {year} {2010}{\natexlab{a}})},\ \Eprint
  {https://arxiv.org/abs/1004.0476} {arXiv:1004.0476 [hep-th]} \BibitemShut
  {NoStop}%
\bibitem [{\citenamefont {Bern}\ \emph
  {et~al.}(2010{\natexlab{b}})\citenamefont {Bern}, \citenamefont {Dennen},
  \citenamefont {Huang},\ and\ \citenamefont {Kiermaier}}]{Bern:2010yg}%
  \BibitemOpen
  \bibfield  {author} {\bibinfo {author} {\bibfnamefont {Z.}~\bibnamefont
  {Bern}}, \bibinfo {author} {\bibfnamefont {T.}~\bibnamefont {Dennen}},
  \bibinfo {author} {\bibfnamefont {Y.-t.}\ \bibnamefont {Huang}},\ and\
  \bibinfo {author} {\bibfnamefont {M.}~\bibnamefont {Kiermaier}},\ }\bibfield
  {title} {\bibinfo {title} {Gravity as the square of gauge theory},\ }\href
  {https://doi.org/10.1103/PhysRevD.82.065003} {\bibfield  {journal} {\bibinfo
  {journal} {Phys. Rev. D}\ }\textbf {\bibinfo {volume} {82}},\ \bibinfo
  {pages} {065003} (\bibinfo {year} {2010}{\natexlab{b}})},\ \Eprint
  {https://arxiv.org/abs/1004.0693} {arXiv:1004.0693 [hep-th]} \BibitemShut
  {NoStop}%
\bibitem [{\citenamefont {Stieberger}(2009)}]{Stieberger:2009hq}%
  \BibitemOpen
  \bibfield  {author} {\bibinfo {author} {\bibfnamefont {S.}~\bibnamefont
  {Stieberger}},\ }\bibfield  {title} {\bibinfo {title} {{Open \& closed
  vs.~pure open string disk amplitudes}},\ }\href@noop {} {\  (\bibinfo {year}
  {2009})},\ \Eprint {https://arxiv.org/abs/0907.2211} {arXiv:0907.2211
  [hep-th]} \BibitemShut {NoStop}%
\bibitem [{\citenamefont {Bjerrum-Bohr}\ \emph {et~al.}(2009)\citenamefont
  {Bjerrum-Bohr}, \citenamefont {Damgaard},\ and\ \citenamefont
  {Vanhove}}]{BjerrumBohr:2009rd}%
  \BibitemOpen
  \bibfield  {author} {\bibinfo {author} {\bibfnamefont {N.~E.~J.}\
  \bibnamefont {Bjerrum-Bohr}}, \bibinfo {author} {\bibfnamefont {P.~H.}\
  \bibnamefont {Damgaard}},\ and\ \bibinfo {author} {\bibfnamefont
  {P.}~\bibnamefont {Vanhove}},\ }\bibfield  {title} {\bibinfo {title} {Minimal
  basis for gauge theory amplitudes},\ }\href
  {https://doi.org/10.1103/PhysRevLett.103.161602} {\bibfield  {journal}
  {\bibinfo  {journal} {Phys. Rev. Lett.}\ }\textbf {\bibinfo {volume} {103}},\
  \bibinfo {pages} {161602} (\bibinfo {year} {2009})},\ \Eprint
  {https://arxiv.org/abs/0907.1425} {arXiv:0907.1425 [hep-th]} \BibitemShut
  {NoStop}%
\bibitem [{\citenamefont {Jia}\ \emph {et~al.}(2010)\citenamefont {Jia},
  \citenamefont {Huang},\ and\ \citenamefont {Liu}}]{Jia:2010nz}%
  \BibitemOpen
  \bibfield  {author} {\bibinfo {author} {\bibfnamefont {Y.}~\bibnamefont
  {Jia}}, \bibinfo {author} {\bibfnamefont {R.}~\bibnamefont {Huang}},\ and\
  \bibinfo {author} {\bibfnamefont {C.-Y.}\ \bibnamefont {Liu}},\ }\bibfield
  {title} {\bibinfo {title} {${U}(1)$-decoupling, {K}leiss--{K}uijf and
  {B}ern--{C}arrasco--{J}ohansson relations in $\mathcal{N}=4$ super
  {Y}ang--{M}ills},\ }\href {https://doi.org/10.1103/PhysRevD.82.065001}
  {\bibfield  {journal} {\bibinfo  {journal} {Phys. Rev. D}\ }\textbf {\bibinfo
  {volume} {82}},\ \bibinfo {pages} {065001} (\bibinfo {year} {2010})},\
  \Eprint {https://arxiv.org/abs/1005.1821} {arXiv:1005.1821 [hep-th]}
  \BibitemShut {NoStop}%
\bibitem [{\citenamefont {Bjerrum-Bohr}\ \emph {et~al.}(2011)\citenamefont
  {Bjerrum-Bohr}, \citenamefont {Damgaard}, \citenamefont {S{\o{}}ndergaard},\
  and\ \citenamefont {Vanhove}}]{BjerrumBohr:2010hn}%
  \BibitemOpen
  \bibfield  {author} {\bibinfo {author} {\bibfnamefont {N.~E.~J.}\
  \bibnamefont {Bjerrum-Bohr}}, \bibinfo {author} {\bibfnamefont {P.~H.}\
  \bibnamefont {Damgaard}}, \bibinfo {author} {\bibfnamefont {T.}~\bibnamefont
  {S{\o{}}ndergaard}},\ and\ \bibinfo {author} {\bibfnamefont {P.}~\bibnamefont
  {Vanhove}},\ }\bibfield  {title} {\bibinfo {title} {The momentum kernel of
  gauge and gravity theories},\ }\href
  {https://doi.org/10.1007/JHEP01(2011)001} {\bibfield  {journal} {\bibinfo
  {journal} {JHEP}\ }\textbf {\bibinfo {volume} {1101}},\ \bibinfo {pages}
  {001}},\ \Eprint {https://arxiv.org/abs/1010.3933} {arXiv:1010.3933 [hep-th]}
  \BibitemShut {NoStop}%
\bibitem [{\citenamefont {Feng}\ \emph {et~al.}(2011)\citenamefont {Feng},
  \citenamefont {Huang},\ and\ \citenamefont {Jia}}]{Feng:2010my}%
  \BibitemOpen
  \bibfield  {author} {\bibinfo {author} {\bibfnamefont {B.}~\bibnamefont
  {Feng}}, \bibinfo {author} {\bibfnamefont {R.}~\bibnamefont {Huang}},\ and\
  \bibinfo {author} {\bibfnamefont {Y.}~\bibnamefont {Jia}},\ }\bibfield
  {title} {\bibinfo {title} {Gauge amplitude identities by on-shell recursion
  relation in {S}-matrix program},\ }\href
  {https://doi.org/10.1016/j.physletb.2010.11.011} {\bibfield  {journal}
  {\bibinfo  {journal} {Phys. Lett. B}\ }\textbf {\bibinfo {volume} {695}},\
  \bibinfo {pages} {350} (\bibinfo {year} {2011})},\ \Eprint
  {https://arxiv.org/abs/1004.3417} {arXiv:1004.3417 [hep-th]} \BibitemShut
  {NoStop}%
\bibitem [{\citenamefont {Chen}\ \emph {et~al.}(2011)\citenamefont {Chen},
  \citenamefont {Du},\ and\ \citenamefont {Feng}}]{Chen:2011jxa}%
  \BibitemOpen
  \bibfield  {author} {\bibinfo {author} {\bibfnamefont {Y.-X.}\ \bibnamefont
  {Chen}}, \bibinfo {author} {\bibfnamefont {Y.-J.}\ \bibnamefont {Du}},\ and\
  \bibinfo {author} {\bibfnamefont {B.}~\bibnamefont {Feng}},\ }\bibfield
  {title} {\bibinfo {title} {A proof of the explicit minimal-basis expansion of
  tree amplitudes in gauge field theory},\ }\href
  {https://doi.org/10.1007/JHEP02(2011)112} {\bibfield  {journal} {\bibinfo
  {journal} {JHEP}\ }\textbf {\bibinfo {volume} {1102}},\ \bibinfo {pages}
  {112}},\ \Eprint {https://arxiv.org/abs/1101.0009} {arXiv:1101.0009 [hep-th]}
  \BibitemShut {NoStop}%
\bibitem [{\citenamefont {Mafra}\ \emph {et~al.}(2011)\citenamefont {Mafra},
  \citenamefont {Schlotterer},\ and\ \citenamefont
  {Stieberger}}]{Mafra:2011kj}%
  \BibitemOpen
  \bibfield  {author} {\bibinfo {author} {\bibfnamefont {C.~R.}\ \bibnamefont
  {Mafra}}, \bibinfo {author} {\bibfnamefont {O.}~\bibnamefont {Schlotterer}},\
  and\ \bibinfo {author} {\bibfnamefont {S.}~\bibnamefont {Stieberger}},\
  }\bibfield  {title} {\bibinfo {title} {Explicit {B}{C}{J} numerators from
  pure spinors},\ }\href {https://doi.org/10.1007/JHEP07(2011)092} {\bibfield
  {journal} {\bibinfo  {journal} {JHEP}\ }\textbf {\bibinfo {volume} {1107}},\
  \bibinfo {pages} {092}},\ \Eprint {https://arxiv.org/abs/1104.5224}
  {arXiv:1104.5224 [hep-th]} \BibitemShut {NoStop}%
\bibitem [{\citenamefont {Du}\ and\ \citenamefont {Fu}(2016)}]{Du:2016tbc}%
  \BibitemOpen
  \bibfield  {author} {\bibinfo {author} {\bibfnamefont {Y.-J.}\ \bibnamefont
  {Du}}\ and\ \bibinfo {author} {\bibfnamefont {C.-H.}\ \bibnamefont {Fu}},\
  }\bibfield  {title} {\bibinfo {title} {Explicit {B}{C}{J} numerators of
  nonlinear sigma model},\ }\href {https://doi.org/10.1007/JHEP09(2016)174}
  {\bibfield  {journal} {\bibinfo  {journal} {JHEP}\ }\textbf {\bibinfo
  {volume} {1609}},\ \bibinfo {pages} {174}},\ \Eprint
  {https://arxiv.org/abs/1606.05846} {arXiv:1606.05846 [hep-th]} \BibitemShut
  {NoStop}%
\bibitem [{\citenamefont {Mizera}(2020)}]{Mizera:2019blq}%
  \BibitemOpen
  \bibfield  {author} {\bibinfo {author} {\bibfnamefont {S.}~\bibnamefont
  {Mizera}},\ }\bibfield  {title} {\bibinfo {title} {Kinematic {J}acobi
  identity is a residue theorem: geometry of color-kinematics duality for gauge
  and gravity amplitudes},\ }\href
  {https://doi.org/10.1103/PhysRevLett.124.141601} {\bibfield  {journal}
  {\bibinfo  {journal} {Phys. Rev. Lett.}\ }\textbf {\bibinfo {volume} {124}},\
  \bibinfo {pages} {141601} (\bibinfo {year} {2020})},\ \Eprint
  {https://arxiv.org/abs/1912.03397} {arXiv:1912.03397 [hep-th]} \BibitemShut
  {NoStop}%
\bibitem [{\citenamefont {Carrasco}(2016)}]{Carrasco:2015iwa}%
  \BibitemOpen
  \bibfield  {author} {\bibinfo {author} {\bibfnamefont {J.~J.~M.}\
  \bibnamefont {Carrasco}},\ }\bibfield  {title} {\bibinfo {title} {Gauge and
  gravity amplitude relations},\ } \bibinfo {note} {in: ``Proceedings, Theoretical
  Advanced Study Institute in Elementary Particle Physics: Journeys Through the
  Precision Frontier: Amplitudes for Colliders (TASI 2014),'' Boulder,
  Colorado, June 2-27, 2014},\ \Eprint {https://arxiv.org/abs/1506.00974}
  {arXiv:1506.00974 [hep-th]} \BibitemShut {NoStop}%
\bibitem [{\citenamefont {Bern}\ \emph {et~al.}(2019)\citenamefont {Bern},
  \citenamefont {Carrasco}, \citenamefont {Chiodaroli}, \citenamefont
  {Johansson},\ and\ \citenamefont {Roiban}}]{Bern:2019prr}%
  \BibitemOpen
  \bibfield  {author} {\bibinfo {author} {\bibfnamefont {Z.}~\bibnamefont
  {Bern}}, \bibinfo {author} {\bibfnamefont {J.~J.}\ \bibnamefont {Carrasco}},
  \bibinfo {author} {\bibfnamefont {M.}~\bibnamefont {Chiodaroli}}, \bibinfo
  {author} {\bibfnamefont {H.}~\bibnamefont {Johansson}},\ and\ \bibinfo
  {author} {\bibfnamefont {R.}~\bibnamefont {Roiban}},\ }\bibfield  {title}
  {\bibinfo {title} {The duality between color and kinematics and its
  applications},\ }\href@noop {} {\  (\bibinfo {year} {2019})},\ \Eprint
  {https://arxiv.org/abs/1909.01358} {arXiv:1909.01358 [hep-th]} \BibitemShut
  {NoStop}%
\bibitem [{\citenamefont {Borsten}(2020)}]{Borsten:2020bgv}%
  \BibitemOpen
  \bibfield  {author} {\bibinfo {author} {\bibfnamefont {L.}~\bibnamefont
  {Borsten}},\ }\bibfield  {title} {\bibinfo {title} {Gravity as the square of
  gauge theory: a review},\ }\href {https://doi.org/10.1007/s40766-020-00003-6}
  {\bibfield  {journal} {\bibinfo  {journal} {Riv. Nuovo Cim.}\ }\textbf
  {\bibinfo {volume} {43}},\ \bibinfo {pages} {97} (\bibinfo {year}
  {2020})}\BibitemShut {NoStop}%
\bibitem [{\citenamefont {Bern}\ \emph
  {et~al.}(2010{\natexlab{c}})\citenamefont {Bern}, \citenamefont {Carrasco},
  \citenamefont {Dixon}, \citenamefont {Johansson},\ and\ \citenamefont
  {Roiban}}]{Bern:2010tq}%
  \BibitemOpen
  \bibfield  {author} {\bibinfo {author} {\bibfnamefont {Z.}~\bibnamefont
  {Bern}}, \bibinfo {author} {\bibfnamefont {J.~J.~M.}\ \bibnamefont
  {Carrasco}}, \bibinfo {author} {\bibfnamefont {L.~J.}\ \bibnamefont {Dixon}},
  \bibinfo {author} {\bibfnamefont {H.}~\bibnamefont {Johansson}},\ and\
  \bibinfo {author} {\bibfnamefont {R.}~\bibnamefont {Roiban}},\ }\bibfield
  {title} {\bibinfo {title} {The complete four-loop four-point amplitude in
  $\mathcal{N}=4$ super-{Y}ang--{M}ills theory},\ }\href
  {https://doi.org/10.1103/PhysRevD.82.125040} {\bibfield  {journal} {\bibinfo
  {journal} {Phys. Rev. D}\ }\textbf {\bibinfo {volume} {82}},\ \bibinfo
  {pages} {125040} (\bibinfo {year} {2010}{\natexlab{c}})},\ \Eprint
  {https://arxiv.org/abs/1008.3327} {arXiv:1008.3327 [hep-th]} \BibitemShut
  {NoStop}%
\bibitem [{\citenamefont {Carrasco}\ and\ \citenamefont
  {Johansson}(2012)}]{Carrasco:2011mn}%
  \BibitemOpen
  \bibfield  {author} {\bibinfo {author} {\bibfnamefont {J.~J.~M.}\
  \bibnamefont {Carrasco}}\ and\ \bibinfo {author} {\bibfnamefont
  {H.}~\bibnamefont {Johansson}},\ }\bibfield  {title} {\bibinfo {title}
  {Five-point amplitudes in $\mathcal{N}=4$ {S}uper-{Y}ang--{M}ills theory and
  $\mathcal{N}=8$ supergravity},\ }\href
  {https://doi.org/10.1103/PhysRevD.85.025006} {\bibfield  {journal} {\bibinfo
  {journal} {Phys. Rev. D}\ }\textbf {\bibinfo {volume} {85}},\ \bibinfo
  {pages} {025006} (\bibinfo {year} {2012})},\ \Eprint
  {https://arxiv.org/abs/1106.4711} {arXiv:1106.4711 [hep-th]} \BibitemShut
  {NoStop}%
\bibitem [{\citenamefont {Bern}\ \emph {et~al.}(2011)\citenamefont {Bern},
  \citenamefont {Boucher-Veronneau},\ and\ \citenamefont
  {Johansson}}]{Bern:2011rj}%
  \BibitemOpen
  \bibfield  {author} {\bibinfo {author} {\bibfnamefont {Z.}~\bibnamefont
  {Bern}}, \bibinfo {author} {\bibfnamefont {C.}~\bibnamefont
  {Boucher-Veronneau}},\ and\ \bibinfo {author} {\bibfnamefont
  {H.}~\bibnamefont {Johansson}},\ }\bibfield  {title} {\bibinfo {title}
  {$\mathcal{N}\leq 4$ supergravity amplitudes from gauge theory at one loop},\
  }\href {https://doi.org/10.1103/PhysRevD.84.105035} {\bibfield  {journal}
  {\bibinfo  {journal} {Phys. Rev. D}\ }\textbf {\bibinfo {volume} {84}},\
  \bibinfo {pages} {105035} (\bibinfo {year} {2011})},\ \Eprint
  {https://arxiv.org/abs/1107.1935} {arXiv:1107.1935 [hep-th]} \BibitemShut
  {NoStop}%
\bibitem [{\citenamefont {Boucher-Veronneau}\ and\ \citenamefont
  {Dixon}(2011)}]{BoucherVeronneau:2011qv}%
  \BibitemOpen
  \bibfield  {author} {\bibinfo {author} {\bibfnamefont {C.}~\bibnamefont
  {Boucher-Veronneau}}\ and\ \bibinfo {author} {\bibfnamefont {L.~J.}\
  \bibnamefont {Dixon}},\ }\bibfield  {title} {\bibinfo {title}
  {$\mathcal{N}\leq 4$ supergravity amplitudes from gauge theory at two
  loops},\ }\href {https://doi.org/10.1007/JHEP12(2011)046} {\bibfield
  {journal} {\bibinfo  {journal} {JHEP}\ }\textbf {\bibinfo {volume} {1112}},\
  \bibinfo {pages} {046}},\ \Eprint {https://arxiv.org/abs/1110.1132}
  {arXiv:1110.1132 [hep-th]} \BibitemShut {NoStop}%
\bibitem [{\citenamefont {Bern}\ \emph
  {et~al.}(2012{\natexlab{a}})\citenamefont {Bern}, \citenamefont {Davies},
  \citenamefont {Dennen},\ and\ \citenamefont {Huang}}]{Bern:2012cd}%
  \BibitemOpen
  \bibfield  {author} {\bibinfo {author} {\bibfnamefont {Z.}~\bibnamefont
  {Bern}}, \bibinfo {author} {\bibfnamefont {S.}~\bibnamefont {Davies}},
  \bibinfo {author} {\bibfnamefont {T.}~\bibnamefont {Dennen}},\ and\ \bibinfo
  {author} {\bibfnamefont {Y.-t.}\ \bibnamefont {Huang}},\ }\bibfield  {title}
  {\bibinfo {title} {Absence of three-loop four-point ultraviolet divergences
  in $\mathcal{N}=4$ supergravity},\ }\href
  {https://doi.org/10.1103/PhysRevLett.108.201301} {\bibfield  {journal}
  {\bibinfo  {journal} {Phys. Rev. Lett.}\ }\textbf {\bibinfo {volume} {108}},\
  \bibinfo {pages} {201301} (\bibinfo {year} {2012}{\natexlab{a}})},\ \Eprint
  {https://arxiv.org/abs/1202.3423} {arXiv:1202.3423 [hep-th]} \BibitemShut
  {NoStop}%
\bibitem [{\citenamefont {Bern}\ \emph
  {et~al.}(2012{\natexlab{b}})\citenamefont {Bern}, \citenamefont {Davies},
  \citenamefont {Dennen},\ and\ \citenamefont {Huang}}]{Bern:2012gh}%
  \BibitemOpen
  \bibfield  {author} {\bibinfo {author} {\bibfnamefont {Z.}~\bibnamefont
  {Bern}}, \bibinfo {author} {\bibfnamefont {S.}~\bibnamefont {Davies}},
  \bibinfo {author} {\bibfnamefont {T.}~\bibnamefont {Dennen}},\ and\ \bibinfo
  {author} {\bibfnamefont {Y.-t.}\ \bibnamefont {Huang}},\ }\bibfield  {title}
  {\bibinfo {title} {Ultraviolet cancellations in half-maximal supergravity as
  a consequence of the double-copy structure},\ }\href
  {https://doi.org/10.1103/PhysRevD.86.105014} {\bibfield  {journal} {\bibinfo
  {journal} {Phys. Rev. D}\ }\textbf {\bibinfo {volume} {86}},\ \bibinfo
  {pages} {105014} (\bibinfo {year} {2012}{\natexlab{b}})},\ \Eprint
  {https://arxiv.org/abs/1209.2472} {arXiv:1209.2472 [hep-th]} \BibitemShut
  {NoStop}%
\bibitem [{\citenamefont {Oxburgh}\ and\ \citenamefont
  {White}(2013)}]{Oxburgh:2012zr}%
  \BibitemOpen
  \bibfield  {author} {\bibinfo {author} {\bibfnamefont {S.}~\bibnamefont
  {Oxburgh}}\ and\ \bibinfo {author} {\bibfnamefont {C.~D.}\ \bibnamefont
  {White}},\ }\bibfield  {title} {\bibinfo {title} {{B}{C}{J} duality and the
  double copy in the soft limit},\ }\href
  {https://doi.org/10.1007/JHEP02(2013)127} {\bibfield  {journal} {\bibinfo
  {journal} {JHEP}\ }\textbf {\bibinfo {volume} {1302}},\ \bibinfo {pages}
  {127}},\ \Eprint {https://arxiv.org/abs/1210.1110} {arXiv:1210.1110 [hep-th]}
  \BibitemShut {NoStop}%
\bibitem [{\citenamefont {Bern}\ \emph
  {et~al.}(2012{\natexlab{c}})\citenamefont {Bern}, \citenamefont {Carrasco},
  \citenamefont {Dixon}, \citenamefont {Johansson},\ and\ \citenamefont
  {Roiban}}]{Bern:2012uf}%
  \BibitemOpen
  \bibfield  {author} {\bibinfo {author} {\bibfnamefont {Z.}~\bibnamefont
  {Bern}}, \bibinfo {author} {\bibfnamefont {J.~J.~M.}\ \bibnamefont
  {Carrasco}}, \bibinfo {author} {\bibfnamefont {L.~J.}\ \bibnamefont {Dixon}},
  \bibinfo {author} {\bibfnamefont {H.}~\bibnamefont {Johansson}},\ and\
  \bibinfo {author} {\bibfnamefont {R.}~\bibnamefont {Roiban}},\ }\bibfield
  {title} {\bibinfo {title} {Simplifying multiloop integrands and ultraviolet
  divergences of gauge theory and gravity amplitudes},\ }\href
  {https://doi.org/10.1103/PhysRevD.85.105014} {\bibfield  {journal} {\bibinfo
  {journal} {Phys. Rev. D}\ }\textbf {\bibinfo {volume} {85}},\ \bibinfo
  {pages} {105014} (\bibinfo {year} {2012}{\natexlab{c}})},\ \Eprint
  {https://arxiv.org/abs/1201.5366} {arXiv:1201.5366 [hep-th]} \BibitemShut
  {NoStop}%
\bibitem [{\citenamefont {Du}\ and\ \citenamefont {Luo}(2013)}]{Du:2012mt}%
  \BibitemOpen
  \bibfield  {author} {\bibinfo {author} {\bibfnamefont {Y.-J.}\ \bibnamefont
  {Du}}\ and\ \bibinfo {author} {\bibfnamefont {H.}~\bibnamefont {Luo}},\
  }\bibfield  {title} {\bibinfo {title} {On general {B}{C}{J} relation at
  one-loop level in {Y}ang--{M}ills theory},\ }\href
  {https://doi.org/10.1007/JHEP01(2013)129} {\bibfield  {journal} {\bibinfo
  {journal} {JHEP}\ }\textbf {\bibinfo {volume} {1301}},\ \bibinfo {pages}
  {129}},\ \Eprint {https://arxiv.org/abs/1207.4549} {arXiv:1207.4549 [hep-th]}
  \BibitemShut {NoStop}%
\bibitem [{\citenamefont {Yuan}(2013)}]{Yuan:2012rg}%
  \BibitemOpen
  \bibfield  {author} {\bibinfo {author} {\bibfnamefont {E.~Y.}\ \bibnamefont
  {Yuan}},\ }\bibfield  {title} {\bibinfo {title} {Virtual color-kinematics
  duality: 6-pt 1-loop {M}{H}{V} amplitudes},\ }\href
  {https://doi.org/10.1007/JHEP05(2013)070} {\bibfield  {journal} {\bibinfo
  {journal} {JHEP}\ }\textbf {\bibinfo {volume} {1305}},\ \bibinfo {pages}
  {070}},\ \Eprint {https://arxiv.org/abs/1210.1816} {arXiv:1210.1816 [hep-th]}
  \BibitemShut {NoStop}%
\bibitem [{\citenamefont {Bern}\ \emph
  {et~al.}(2013{\natexlab{a}})\citenamefont {Bern}, \citenamefont {Davies},
  \citenamefont {Dennen}, \citenamefont {Smirnov},\ and\ \citenamefont
  {Smirnov}}]{Bern:2013uka}%
  \BibitemOpen
  \bibfield  {author} {\bibinfo {author} {\bibfnamefont {Z.}~\bibnamefont
  {Bern}}, \bibinfo {author} {\bibfnamefont {S.}~\bibnamefont {Davies}},
  \bibinfo {author} {\bibfnamefont {T.}~\bibnamefont {Dennen}}, \bibinfo
  {author} {\bibfnamefont {A.~V.}\ \bibnamefont {Smirnov}},\ and\ \bibinfo
  {author} {\bibfnamefont {V.~A.}\ \bibnamefont {Smirnov}},\ }\bibfield
  {title} {\bibinfo {title} {Ultraviolet properties of $\mathcal{N}=4$
  supergravity at four loops},\ }\href
  {https://doi.org/10.1103/PhysRevLett.111.231302} {\bibfield  {journal}
  {\bibinfo  {journal} {Phys. Rev. Lett.}\ }\textbf {\bibinfo {volume} {111}},\
  \bibinfo {pages} {231302} (\bibinfo {year} {2013}{\natexlab{a}})},\ \Eprint
  {https://arxiv.org/abs/1309.2498} {arXiv:1309.2498 [hep-th]} \BibitemShut
  {NoStop}%
\bibitem [{\citenamefont {Boels}\ \emph {et~al.}(2013)\citenamefont {Boels},
  \citenamefont {Isermann}, \citenamefont {Monteiro},\ and\ \citenamefont
  {O'Connell}}]{Boels:2013bi}%
  \BibitemOpen
  \bibfield  {author} {\bibinfo {author} {\bibfnamefont {R.~H.}\ \bibnamefont
  {Boels}}, \bibinfo {author} {\bibfnamefont {R.~S.}\ \bibnamefont {Isermann}},
  \bibinfo {author} {\bibfnamefont {R.}~\bibnamefont {Monteiro}},\ and\
  \bibinfo {author} {\bibfnamefont {D.}~\bibnamefont {O'Connell}},\ }\bibfield
  {title} {\bibinfo {title} {Colour-kinematics duality for one-loop rational
  amplitudes},\ }\href {https://doi.org/10.1007/JHEP04(2013)107} {\bibfield
  {journal} {\bibinfo  {journal} {JHEP}\ }\textbf {\bibinfo {volume} {1304}},\
  \bibinfo {pages} {107}},\ \Eprint {https://arxiv.org/abs/1301.4165}
  {arXiv:1301.4165 [hep-th]} \BibitemShut {NoStop}%
\bibitem [{\citenamefont {Bern}\ \emph {et~al.}(2015)\citenamefont {Bern},
  \citenamefont {Davies}, \citenamefont {Dennen}, \citenamefont {Huang},\ and\
  \citenamefont {Nohle}}]{Bern:2013yya}%
  \BibitemOpen
  \bibfield  {author} {\bibinfo {author} {\bibfnamefont {Z.}~\bibnamefont
  {Bern}}, \bibinfo {author} {\bibfnamefont {S.}~\bibnamefont {Davies}},
  \bibinfo {author} {\bibfnamefont {T.}~\bibnamefont {Dennen}}, \bibinfo
  {author} {\bibfnamefont {Y.-t.}\ \bibnamefont {Huang}},\ and\ \bibinfo
  {author} {\bibfnamefont {J.}~\bibnamefont {Nohle}},\ }\bibfield  {title}
  {\bibinfo {title} {Color-kinematics duality for pure {Y}ang--{M}ills and
  gravity at one and two loops},\ }\href
  {https://doi.org/10.1103/PhysRevD.92.045041} {\bibfield  {journal} {\bibinfo
  {journal} {Phys. Rev. D}\ }\textbf {\bibinfo {volume} {92}},\ \bibinfo
  {pages} {045041} (\bibinfo {year} {2015})},\ \Eprint
  {https://arxiv.org/abs/1303.6605} {arXiv:1303.6605 [hep-th]} \BibitemShut
  {NoStop}%
\bibitem [{\citenamefont {Bern}\ \emph
  {et~al.}(2013{\natexlab{b}})\citenamefont {Bern}, \citenamefont {Davies},\
  and\ \citenamefont {Dennen}}]{Bern:2013qca}%
  \BibitemOpen
  \bibfield  {author} {\bibinfo {author} {\bibfnamefont {Z.}~\bibnamefont
  {Bern}}, \bibinfo {author} {\bibfnamefont {S.}~\bibnamefont {Davies}},\ and\
  \bibinfo {author} {\bibfnamefont {T.}~\bibnamefont {Dennen}},\ }\bibfield
  {title} {\bibinfo {title} {The ultraviolet structure of half-maximal
  supergravity with matter multiplets at two and three loops},\ }\href
  {https://doi.org/10.1103/PhysRevD.88.065007} {\bibfield  {journal} {\bibinfo
  {journal} {Phys. Rev. D}\ }\textbf {\bibinfo {volume} {88}},\ \bibinfo
  {pages} {065007} (\bibinfo {year} {2013}{\natexlab{b}})},\ \Eprint
  {https://arxiv.org/abs/1305.4876} {arXiv:1305.4876 [hep-th]} \BibitemShut
  {NoStop}%
\bibitem [{\citenamefont {Bern}\ \emph
  {et~al.}(2014{\natexlab{a}})\citenamefont {Bern}, \citenamefont {Davies},\
  and\ \citenamefont {Dennen}}]{Bern:2014lha}%
  \BibitemOpen
  \bibfield  {author} {\bibinfo {author} {\bibfnamefont {Z.}~\bibnamefont
  {Bern}}, \bibinfo {author} {\bibfnamefont {S.}~\bibnamefont {Davies}},\ and\
  \bibinfo {author} {\bibfnamefont {T.}~\bibnamefont {Dennen}},\ }\bibfield
  {title} {\bibinfo {title} {The ultraviolet critical dimension of half-maximal
  supergravity at three loops},\ }\href@noop {} {\  (\bibinfo {year}
  {2014}{\natexlab{a}})},\ \Eprint {https://arxiv.org/abs/1412.2441}
  {arXiv:1412.2441 [hep-th]} \BibitemShut {NoStop}%
\bibitem [{\citenamefont {Bern}\ \emph
  {et~al.}(2014{\natexlab{b}})\citenamefont {Bern}, \citenamefont {Davies},\
  and\ \citenamefont {Dennen}}]{Bern:2014sna}%
  \BibitemOpen
  \bibfield  {author} {\bibinfo {author} {\bibfnamefont {Z.}~\bibnamefont
  {Bern}}, \bibinfo {author} {\bibfnamefont {S.}~\bibnamefont {Davies}},\ and\
  \bibinfo {author} {\bibfnamefont {T.}~\bibnamefont {Dennen}},\ }\bibfield
  {title} {\bibinfo {title} {Enhanced ultraviolet cancellations in
  $\mathcal{N}=5$ supergravity at four loop},\ }\href
  {https://doi.org/10.1103/PhysRevD.90.105011} {\bibfield  {journal} {\bibinfo
  {journal} {Phys. Rev. D}\ }\textbf {\bibinfo {volume} {90}},\ \bibinfo
  {pages} {105011} (\bibinfo {year} {2014}{\natexlab{b}})},\ \Eprint
  {https://arxiv.org/abs/1409.3089} {arXiv:1409.3089 [hep-th]} \BibitemShut
  {NoStop}%
\bibitem [{\citenamefont {Mafra}\ and\ \citenamefont
  {Schlotterer}(2015{\natexlab{a}})}]{Mafra:2014gja}%
  \BibitemOpen
  \bibfield  {author} {\bibinfo {author} {\bibfnamefont {C.~R.}\ \bibnamefont
  {Mafra}}\ and\ \bibinfo {author} {\bibfnamefont {O.}~\bibnamefont
  {Schlotterer}},\ }\bibfield  {title} {\bibinfo {title} {Towards one-loop
  {SYM} amplitudes from the pure spinor {BRST} cohomology},\ }\href
  {https://doi.org/10.1002/prop.201400076} {\bibfield  {journal} {\bibinfo
  {journal} {Fortsch. Phys.}\ }\textbf {\bibinfo {volume} {63}},\ \bibinfo
  {pages} {105} (\bibinfo {year} {2015}{\natexlab{a}})},\ \Eprint
  {https://arxiv.org/abs/1410.0668} {arXiv:1410.0668 [hep-th]} \BibitemShut
  {NoStop}%
\bibitem [{\citenamefont {Mafra}\ and\ \citenamefont
  {Schlotterer}(2015{\natexlab{b}})}]{Mafra:2015mja}%
  \BibitemOpen
  \bibfield  {author} {\bibinfo {author} {\bibfnamefont {C.~R.}\ \bibnamefont
  {Mafra}}\ and\ \bibinfo {author} {\bibfnamefont {O.}~\bibnamefont
  {Schlotterer}},\ }\bibfield  {title} {\bibinfo {title} {Two-loop five-point
  amplitudes of super {Y}ang--{M}ills and supergravity in pure spinor
  superspace},\ }\href {https://doi.org/10.1007/JHEP10(2015)124} {\bibfield
  {journal} {\bibinfo  {journal} {JHEP}\ }\textbf {\bibinfo {volume} {1510}},\
  \bibinfo {pages} {124}},\ \Eprint {https://arxiv.org/abs/1505.02746}
  {arXiv:1505.02746 [hep-th]} \BibitemShut {NoStop}%
\bibitem [{\citenamefont {Johansson}\ \emph {et~al.}(2017)\citenamefont
  {Johansson}, \citenamefont {Kälin},\ and\ \citenamefont
  {Mogull}}]{Johansson:2017bfl}%
  \BibitemOpen
  \bibfield  {author} {\bibinfo {author} {\bibfnamefont {H.}~\bibnamefont
  {Johansson}}, \bibinfo {author} {\bibfnamefont {G.}~\bibnamefont {Kälin}},\
  and\ \bibinfo {author} {\bibfnamefont {G.}~\bibnamefont {Mogull}},\
  }\bibfield  {title} {\bibinfo {title} {Two-loop supersymmetric {Q}{C}{D} and
  half-maximal supergravity amplitudes},\ }\href
  {https://doi.org/10.1007/JHEP09(2017)019} {\bibfield  {journal} {\bibinfo
  {journal} {JHEP}\ }\textbf {\bibinfo {volume} {1709}},\ \bibinfo {pages}
  {019}},\ \Eprint {https://arxiv.org/abs/1706.09381} {arXiv:1706.09381
  [hep-th]} \BibitemShut {NoStop}%
\bibitem [{\citenamefont {Bern}\ \emph
  {et~al.}(2017{\natexlab{a}})\citenamefont {Bern}, \citenamefont {Carrasco},
  \citenamefont {Chen}, \citenamefont {Johansson}, \citenamefont {Roiban},\
  and\ \citenamefont {Zeng}}]{Bern:2017ucb}%
  \BibitemOpen
  \bibfield  {author} {\bibinfo {author} {\bibfnamefont {Z.}~\bibnamefont
  {Bern}}, \bibinfo {author} {\bibfnamefont {J.~J.~M.}\ \bibnamefont
  {Carrasco}}, \bibinfo {author} {\bibfnamefont {W.-M.}\ \bibnamefont {Chen}},
  \bibinfo {author} {\bibfnamefont {H.}~\bibnamefont {Johansson}}, \bibinfo
  {author} {\bibfnamefont {R.}~\bibnamefont {Roiban}},\ and\ \bibinfo {author}
  {\bibfnamefont {M.}~\bibnamefont {Zeng}},\ }\bibfield  {title} {\bibinfo
  {title} {The five-loop four-point integrand of $\mathcal{N}=8$ supergravity
  as a generalized double copy},\ }\href
  {https://doi.org/10.1103/PhysRevD.96.126012} {\bibfield  {journal} {\bibinfo
  {journal} {Phys. Rev. D}\ }\textbf {\bibinfo {volume} {96}},\ \bibinfo
  {pages} {126012} (\bibinfo {year} {2017}{\natexlab{a}})},\ \Eprint
  {https://arxiv.org/abs/1708.06807} {arXiv:1708.06807 [hep-th]} \BibitemShut
  {NoStop}%
\bibitem [{\citenamefont {Bern}\ \emph {et~al.}(2018)\citenamefont {Bern},
  \citenamefont {Carrasco}, \citenamefont {Chen}, \citenamefont {Edison},
  \citenamefont {Johansson}, \citenamefont {Parra-Martinez}, \citenamefont
  {Roiban},\ and\ \citenamefont {Zeng}}]{Bern:2018jmv}%
  \BibitemOpen
  \bibfield  {author} {\bibinfo {author} {\bibfnamefont {Z.}~\bibnamefont
  {Bern}}, \bibinfo {author} {\bibfnamefont {J.~J.}\ \bibnamefont {Carrasco}},
  \bibinfo {author} {\bibfnamefont {W.-M.}\ \bibnamefont {Chen}}, \bibinfo
  {author} {\bibfnamefont {A.}~\bibnamefont {Edison}}, \bibinfo {author}
  {\bibfnamefont {H.}~\bibnamefont {Johansson}}, \bibinfo {author}
  {\bibfnamefont {J.}~\bibnamefont {Parra-Martinez}}, \bibinfo {author}
  {\bibfnamefont {R.}~\bibnamefont {Roiban}},\ and\ \bibinfo {author}
  {\bibfnamefont {M.}~\bibnamefont {Zeng}},\ }\bibfield  {title} {\bibinfo
  {title} {Ultraviolet properties of $\mathcal{N}=8$ supergravity at five
  loops},\ }\href {https://doi.org/10.1103/PhysRevD.98.086021} {\bibfield
  {journal} {\bibinfo  {journal} {Phys. Rev. D}\ }\textbf {\bibinfo {volume}
  {98}},\ \bibinfo {pages} {086021} (\bibinfo {year} {2018})},\ \Eprint
  {https://arxiv.org/abs/1804.09311} {arXiv:1804.09311 [hep-th]} \BibitemShut
  {NoStop}%
\bibitem [{\citenamefont {Monteiro}\ and\ \citenamefont
  {O'Connell}(2011)}]{Monteiro:2011pc}%
  \BibitemOpen
  \bibfield  {author} {\bibinfo {author} {\bibfnamefont {R.}~\bibnamefont
  {Monteiro}}\ and\ \bibinfo {author} {\bibfnamefont {D.}~\bibnamefont
  {O'Connell}},\ }\bibfield  {title} {\bibinfo {title} {The kinematic algebra
  from the self-dual sector},\ }\href {https://doi.org/10.1007/JHEP07(2011)007}
  {\bibfield  {journal} {\bibinfo  {journal} {JHEP}\ }\textbf {\bibinfo
  {volume} {1107}},\ \bibinfo {pages} {007}},\ \Eprint
  {https://arxiv.org/abs/1105.2565} {arXiv:1105.2565 [hep-th]} \BibitemShut
  {NoStop}%
\bibitem [{\citenamefont {Bjerrum-Bohr}\ \emph {et~al.}(2012)\citenamefont
  {Bjerrum-Bohr}, \citenamefont {Damgaard}, \citenamefont {Monteiro},\ and\
  \citenamefont {O'Connell}}]{BjerrumBohr:2012mg}%
  \BibitemOpen
  \bibfield  {author} {\bibinfo {author} {\bibfnamefont {N.~E.~J.}\
  \bibnamefont {Bjerrum-Bohr}}, \bibinfo {author} {\bibfnamefont {P.~H.}\
  \bibnamefont {Damgaard}}, \bibinfo {author} {\bibfnamefont {R.}~\bibnamefont
  {Monteiro}},\ and\ \bibinfo {author} {\bibfnamefont {D.}~\bibnamefont
  {O'Connell}},\ }\bibfield  {title} {\bibinfo {title} {Algebras for
  amplitudes},\ }\href {https://doi.org/10.1007/JHEP06(2012)061} {\bibfield
  {journal} {\bibinfo  {journal} {JHEP}\ }\textbf {\bibinfo {volume} {1206}},\
  \bibinfo {pages} {061}},\ \Eprint {https://arxiv.org/abs/1203.0944}
  {arXiv:1203.0944 [hep-th]} \BibitemShut {NoStop}%
\bibitem [{\citenamefont {Tolotti}\ and\ \citenamefont
  {Weinzierl}(2013)}]{Tolotti:2013caa}%
  \BibitemOpen
  \bibfield  {author} {\bibinfo {author} {\bibfnamefont {M.}~\bibnamefont
  {Tolotti}}\ and\ \bibinfo {author} {\bibfnamefont {S.}~\bibnamefont
  {Weinzierl}},\ }\bibfield  {title} {\bibinfo {title} {Construction of an
  effective {Y}ang--{M}ills {L}agrangian with manifest {B}{C}{J} duality},\
  }\href {https://doi.org/10.1007/JHEP07(2013)111} {\bibfield  {journal}
  {\bibinfo  {journal} {JHEP}\ }\textbf {\bibinfo {volume} {1307}},\ \bibinfo
  {pages} {111}},\ \Eprint {https://arxiv.org/abs/1306.2975} {arXiv:1306.2975
  [hep-th]} \BibitemShut {NoStop}%
\bibitem [{\citenamefont {Monteiro}\ and\ \citenamefont
  {O'Connell}(2014)}]{Monteiro:2013rya}%
  \BibitemOpen
  \bibfield  {author} {\bibinfo {author} {\bibfnamefont {R.}~\bibnamefont
  {Monteiro}}\ and\ \bibinfo {author} {\bibfnamefont {D.}~\bibnamefont
  {O'Connell}},\ }\bibfield  {title} {\bibinfo {title} {The kinematic algebras
  from the scattering equations},\ }\href
  {https://doi.org/10.1007/JHEP03(2014)110} {\bibfield  {journal} {\bibinfo
  {journal} {JHEP}\ }\textbf {\bibinfo {volume} {1403}},\ \bibinfo {pages}
  {110}},\ \Eprint {https://arxiv.org/abs/1311.1151} {arXiv:1311.1151 [hep-th]}
  \BibitemShut {NoStop}%
\bibitem [{\citenamefont {Fu}\ and\ \citenamefont
  {Krasnov}(2017)}]{Fu:2016plh}%
  \BibitemOpen
  \bibfield  {author} {\bibinfo {author} {\bibfnamefont {C.-H.}\ \bibnamefont
  {Fu}}\ and\ \bibinfo {author} {\bibfnamefont {K.}~\bibnamefont {Krasnov}},\
  }\bibfield  {title} {\bibinfo {title} {Colour-kinematics duality and the
  {D}rinfeld double of the {L}ie algebra of diffeomorphisms},\ }\href
  {https://doi.org/10.1007/JHEP01(2017)075} {\bibfield  {journal} {\bibinfo
  {journal} {JHEP}\ }\textbf {\bibinfo {volume} {1701}},\ \bibinfo {pages}
  {075}},\ \Eprint {https://arxiv.org/abs/1603.02033} {arXiv:1603.02033
  [hep-th]} \BibitemShut {NoStop}%
\bibitem [{\citenamefont {Cheung}\ and\ \citenamefont
  {Shen}(2017)}]{Cheung:2016prv}%
  \BibitemOpen
  \bibfield  {author} {\bibinfo {author} {\bibfnamefont {C.}~\bibnamefont
  {Cheung}}\ and\ \bibinfo {author} {\bibfnamefont {C.-H.}\ \bibnamefont
  {Shen}},\ }\bibfield  {title} {\bibinfo {title} {Symmetry for
  flavor-kinematics duality from action},\ }\href
  {https://doi.org/10.1103/PhysRevLett.118.121601} {\bibfield  {journal}
  {\bibinfo  {journal} {Phys. Rev. Lett.}\ }\textbf {\bibinfo {volume} {118}},\
  \bibinfo {pages} {121601} (\bibinfo {year} {2017})},\ \Eprint
  {https://arxiv.org/abs/1612.00868} {arXiv:1612.00868 [hep-th]} \BibitemShut
  {NoStop}%
\bibitem [{\citenamefont {Chen}\ \emph {et~al.}(2019)\citenamefont {Chen},
  \citenamefont {Johansson}, \citenamefont {Teng},\ and\ \citenamefont
  {Wang}}]{Chen:2019ywi}%
  \BibitemOpen
  \bibfield  {author} {\bibinfo {author} {\bibfnamefont {G.}~\bibnamefont
  {Chen}}, \bibinfo {author} {\bibfnamefont {H.}~\bibnamefont {Johansson}},
  \bibinfo {author} {\bibfnamefont {F.}~\bibnamefont {Teng}},\ and\ \bibinfo
  {author} {\bibfnamefont {T.}~\bibnamefont {Wang}},\ }\bibfield  {title}
  {\bibinfo {title} {On the kinematic algebra for {B}{C}{J} numerators beyond
  the {M}{H}{V} sector},\ }\href {https://doi.org/10.1007/JHEP11(2019)055}
  {\bibfield  {journal} {\bibinfo  {journal} {JHEP}\ }\textbf {\bibinfo
  {volume} {2019}},\ \bibinfo {pages} {055}},\ \Eprint
  {https://arxiv.org/abs/1906.10683} {arXiv:1906.10683 [hep-th]} \BibitemShut
  {NoStop}%
\bibitem [{\citenamefont {Borsten}\ and\ \citenamefont
  {Nagy}(2020)}]{Borsten:2020xbt}%
  \BibitemOpen
  \bibfield  {author} {\bibinfo {author} {\bibfnamefont {L.}~\bibnamefont
  {Borsten}}\ and\ \bibinfo {author} {\bibfnamefont {S.}~\bibnamefont {Nagy}},\
  }\bibfield  {title} {\bibinfo {title} {The pure {B}{R}{S}{T}
  {E}instein--{H}ilbert {L}agrangian from the double-copy to cubic order},\
  }\href {https://doi.org/10.1007/JHEP07(2020)093} {\bibfield  {journal}
  {\bibinfo  {journal} {JHEP}\ }\textbf {\bibinfo {volume} {2007}},\ \bibinfo
  {pages} {093}},\ \Eprint {https://arxiv.org/abs/2004.14945} {arXiv:2004.14945
  [hep-th]} \BibitemShut {NoStop}%
\bibitem [{\citenamefont {Kugo}\ and\ \citenamefont
  {Ojima}(1978)}]{Kugo:1977yx}%
  \BibitemOpen
  \bibfield  {author} {\bibinfo {author} {\bibfnamefont {T.}~\bibnamefont
  {Kugo}}\ and\ \bibinfo {author} {\bibfnamefont {I.}~\bibnamefont {Ojima}},\
  }\bibfield  {title} {\bibinfo {title} {{Manifestly covariant canonical
  formulation of {Y}ang--{M}ills field theories. I. General formalism}},\
  }\href {https://doi.org/10.1143/PTP.60.1869} {\bibfield  {journal} {\bibinfo
  {journal} {Prog. Theor. Phys.}\ }\textbf {\bibinfo {volume} {60}},\ \bibinfo
  {pages} {1869} (\bibinfo {year} {1978})}\BibitemShut {NoStop}%
\bibitem [{\citenamefont {Anastasiou}\ \emph {et~al.}(2014)\citenamefont
  {Anastasiou}, \citenamefont {Borsten}, \citenamefont {Duff}, \citenamefont
  {Hughes},\ and\ \citenamefont {Nagy}}]{Anastasiou:2014qba}%
  \BibitemOpen
  \bibfield  {author} {\bibinfo {author} {\bibfnamefont {A.}~\bibnamefont
  {Anastasiou}}, \bibinfo {author} {\bibfnamefont {L.}~\bibnamefont {Borsten}},
  \bibinfo {author} {\bibfnamefont {M.~J.}\ \bibnamefont {Duff}}, \bibinfo
  {author} {\bibfnamefont {L.~J.}\ \bibnamefont {Hughes}},\ and\ \bibinfo
  {author} {\bibfnamefont {S.}~\bibnamefont {Nagy}},\ }\bibfield  {title}
  {\bibinfo {title} {{Y}ang--{M}ills origin of gravitational symmetries},\
  }\href {https://doi.org/10.1103/PhysRevLett.113.231606} {\bibfield  {journal}
  {\bibinfo  {journal} {Phys. Rev. Lett.}\ }\textbf {\bibinfo {volume} {113}},\
  \bibinfo {pages} {231606} (\bibinfo {year} {2014})},\ \Eprint
  {https://arxiv.org/abs/1408.4434} {arXiv:1408.4434 [hep-th]} \BibitemShut
  {NoStop}%
\bibitem [{\citenamefont {Borsten}(2018)}]{Borsten:2017jpt}%
  \BibitemOpen
  \bibfield  {author} {\bibinfo {author} {\bibfnamefont {L.}~\bibnamefont
  {Borsten}},\ }\bibfield  {title} {\bibinfo {title} {{On $D=6$,
  $\mathcal{N}=(2,0)$ and $\mathcal{N}=(4,0)$ theories}},\ }\href
  {https://doi.org/10.1103/PhysRevD.97.066014} {\bibfield  {journal} {\bibinfo
  {journal} {Phys. Rev. D}\ }\textbf {\bibinfo {volume} {97}},\ \bibinfo
  {pages} {066014} (\bibinfo {year} {2018})},\ \Eprint
  {https://arxiv.org/abs/1708.02573} {arXiv:1708.02573 [hep-th]} \BibitemShut
  {NoStop}%
\bibitem [{\citenamefont {Anastasiou}\ \emph {et~al.}(2018)\citenamefont
  {Anastasiou}, \citenamefont {Borsten}, \citenamefont {Duff}, \citenamefont
  {Nagy},\ and\ \citenamefont {Zoccali}}]{Anastasiou:2018rdx}%
  \BibitemOpen
  \bibfield  {author} {\bibinfo {author} {\bibfnamefont {A.}~\bibnamefont
  {Anastasiou}}, \bibinfo {author} {\bibfnamefont {L.}~\bibnamefont {Borsten}},
  \bibinfo {author} {\bibfnamefont {M.~J.}\ \bibnamefont {Duff}}, \bibinfo
  {author} {\bibfnamefont {S.}~\bibnamefont {Nagy}},\ and\ \bibinfo {author}
  {\bibfnamefont {M.}~\bibnamefont {Zoccali}},\ }\bibfield  {title} {\bibinfo
  {title} {Gravity as gauge theory squared: a ghost story},\ }\href
  {https://doi.org/10.1103/PhysRevLett.121.211601} {\bibfield  {journal}
  {\bibinfo  {journal} {Phys. Rev. Lett.}\ }\textbf {\bibinfo {volume} {121}},\
  \bibinfo {pages} {211601} (\bibinfo {year} {2018})},\ \Eprint
  {https://arxiv.org/abs/1807.02486} {arXiv:1807.02486 [hep-th]} \BibitemShut
  {NoStop}%
\bibitem [{\citenamefont {Zoccali}(2018)}]{Zoccali:2018thesis}%
  \BibitemOpen
  \bibfield  {author} {\bibinfo {author} {\bibfnamefont {M.}~\bibnamefont
  {Zoccali}},\ }\emph {\bibinfo {title} {Supergravity as Yang--Mills
  squared}},\ \href {https://doi.org/10.25560/67915} {Ph.D. thesis} (\bibinfo
  {year} {2018})\BibitemShut {NoStop}%
\bibitem [{\citenamefont {Borsten}\ \emph {et~al.}(2020)\citenamefont
  {Borsten}, \citenamefont {Jubb}, \citenamefont {Makwana},\ and\ \citenamefont
  {Nagy}}]{Borsten:2019prq}%
  \BibitemOpen
  \bibfield  {author} {\bibinfo {author} {\bibfnamefont {L.}~\bibnamefont
  {Borsten}}, \bibinfo {author} {\bibfnamefont {I.}~\bibnamefont {Jubb}},
  \bibinfo {author} {\bibfnamefont {V.}~\bibnamefont {Makwana}},\ and\ \bibinfo
  {author} {\bibfnamefont {S.}~\bibnamefont {Nagy}},\ }\bibfield  {title}
  {\bibinfo {title} {Gauge $\times$ gauge on spheres},\ }\href
  {https://doi.org/10.1007/JHEP06(2020)096} {\bibfield  {journal} {\bibinfo
  {journal} {JHEP}\ }\textbf {\bibinfo {volume} {2006}},\ \bibinfo {pages}
  {096}},\ \Eprint {https://arxiv.org/abs/1911.12324} {arXiv:1911.12324
  [hep-th]} \BibitemShut {NoStop}%
\bibitem [{\citenamefont {Luna}\ \emph {et~al.}(2020)\citenamefont {Luna},
  \citenamefont {Nagy},\ and\ \citenamefont {White}}]{Luna:2020adi}%
  \BibitemOpen
  \bibfield  {author} {\bibinfo {author} {\bibfnamefont {A.}~\bibnamefont
  {Luna}}, \bibinfo {author} {\bibfnamefont {S.}~\bibnamefont {Nagy}},\ and\
  \bibinfo {author} {\bibfnamefont {C.}~\bibnamefont {White}},\ }\bibfield
  {title} {\bibinfo {title} {The convolutional double copy: a case study with a
  point},\ }\href {https://doi.org/10.1007/JHEP09(2020)062} {\bibfield
  {journal} {\bibinfo  {journal} {JHEP}\ }\textbf {\bibinfo {volume} {2009}},\
  \bibinfo {pages} {062}},\ \Eprint {https://arxiv.org/abs/2004.11254}
  {arXiv:2004.11254 [hep-th]} \BibitemShut {NoStop}%
\bibitem [{\citenamefont {Borsten}\ \emph {et~al.}()\citenamefont {Borsten},
  \citenamefont {Jur\v{c}o}, \citenamefont {Kim}, \citenamefont {Macrelli},
  \citenamefont {Saemann},\ and\ \citenamefont {Wolf}}]{Borsten:2021hua}%
  \BibitemOpen
  \bibfield  {author} {\bibinfo {author} {\bibfnamefont {L.}~\bibnamefont
  {Borsten}}, \bibinfo {author} {\bibfnamefont {B.}~\bibnamefont {Jur\v{c}o}},
  \bibinfo {author} {\bibfnamefont {H.}~\bibnamefont {Kim}}, \bibinfo {author}
  {\bibfnamefont {T.}~\bibnamefont {Macrelli}}, \bibinfo {author}
  {\bibfnamefont {C.}~\bibnamefont {Saemann}},\ and\ \bibinfo {author}
  {\bibfnamefont {M.}~\bibnamefont {Wolf}},\ }\bibfield  {title} {\bibinfo
  {title} {Double copy from homotopy algebras},\ }\href@noop {} {\ }\Eprint
  {https://arxiv.org/abs/2102.11390} {arXiv:2102.11390 [hep-th]} \BibitemShut
  {NoStop}%
\bibitem [{\citenamefont {Bern}\ \emph
  {et~al.}(2017{\natexlab{b}})\citenamefont {Bern}, \citenamefont {Carrasco},
  \citenamefont {Chen}, \citenamefont {Johansson},\ and\ \citenamefont
  {Roiban}}]{Bern:2017yxu}%
  \BibitemOpen
  \bibfield  {author} {\bibinfo {author} {\bibfnamefont {Z.}~\bibnamefont
  {Bern}}, \bibinfo {author} {\bibfnamefont {J.~J.}\ \bibnamefont {Carrasco}},
  \bibinfo {author} {\bibfnamefont {W.-M.}\ \bibnamefont {Chen}}, \bibinfo
  {author} {\bibfnamefont {H.}~\bibnamefont {Johansson}},\ and\ \bibinfo
  {author} {\bibfnamefont {R.}~\bibnamefont {Roiban}},\ }\bibfield  {title}
  {\bibinfo {title} {Gravity amplitudes as generalized double copies},\ }\href
  {https://doi.org/10.1103/PhysRevLett.118.181602} {\bibfield  {journal}
  {\bibinfo  {journal} {Phys. Rev. Lett.}\ }\textbf {\bibinfo {volume} {118}},\
  \bibinfo {pages} {181602} (\bibinfo {year} {2017}{\natexlab{b}})},\ \Eprint
  {https://arxiv.org/abs/1701.02519} {arXiv:1701.02519 [hep-th]} \BibitemShut
  {NoStop}%
\bibitem [{\citenamefont {Casali}\ \emph {et~al.}(2021)\citenamefont {Casali},
  \citenamefont {Mizera},\ and\ \citenamefont {Tourkine}}]{Casali:2020knc}%
  \BibitemOpen
  \bibfield  {author} {\bibinfo {author} {\bibfnamefont {E.}~\bibnamefont
  {Casali}}, \bibinfo {author} {\bibfnamefont {S.}~\bibnamefont {Mizera}},\
  and\ \bibinfo {author} {\bibfnamefont {P.}~\bibnamefont {Tourkine}},\
  }\bibfield  {title} {\bibinfo {title} {Loop amplitudes monodromy relations
  and color-kinematics duality},\ }\href
  {https://doi.org/10.1007/JHEP03(2021)048} {\bibfield  {journal} {\bibinfo
  {journal} {JHEP}\ }\textbf {\bibinfo {volume} {2021}},\ \bibinfo {pages}
  {048}},\ \Eprint {https://arxiv.org/abs/2005.05329} {arXiv:2005.05329
  [hep-th]} \BibitemShut {NoStop}%
\bibitem [{\citenamefont {Jur\v{c}o}\ \emph {et~al.}(2019)\citenamefont
  {Jur\v{c}o}, \citenamefont {Raspollini}, \citenamefont {Saemann},\ and\
  \citenamefont {Wolf}}]{Jurco:2018sby}%
  \BibitemOpen
  \bibfield  {author} {\bibinfo {author} {\bibfnamefont {B.}~\bibnamefont
  {Jur\v{c}o}}, \bibinfo {author} {\bibfnamefont {L.}~\bibnamefont
  {Raspollini}}, \bibinfo {author} {\bibfnamefont {C.}~\bibnamefont
  {Saemann}},\ and\ \bibinfo {author} {\bibfnamefont {M.}~\bibnamefont
  {Wolf}},\ }\bibfield  {title} {\bibinfo {title} {{$L_\infty$-algebras of
  classical field theories and the Batalin--Vilkovisky formalism}},\ }\href
  {https://doi.org/10.1002/prop.201900025} {\bibfield  {journal} {\bibinfo
  {journal} {Fortsch. Phys.}\ }\textbf {\bibinfo {volume} {67}},\ \bibinfo
  {pages} {1900025} (\bibinfo {year} {2019})},\ \Eprint
  {https://arxiv.org/abs/1809.09899} {arXiv:1809.09899 [hep-th]} \BibitemShut
  {NoStop}%
\bibitem [{\citenamefont {Macrelli}\ \emph {et~al.}(2019)\citenamefont
  {Macrelli}, \citenamefont {Saemann},\ and\ \citenamefont
  {Wolf}}]{Macrelli:2019afx}%
  \BibitemOpen
  \bibfield  {author} {\bibinfo {author} {\bibfnamefont {T.}~\bibnamefont
  {Macrelli}}, \bibinfo {author} {\bibfnamefont {C.}~\bibnamefont {Saemann}},\
  and\ \bibinfo {author} {\bibfnamefont {M.}~\bibnamefont {Wolf}},\ }\bibfield
  {title} {\bibinfo {title} {{Scattering amplitude recursion relations in BV
  quantisable theories}},\ }\href {https://doi.org/10.1103/PhysRevD.100.045017}
  {\bibfield  {journal} {\bibinfo  {journal} {Phys. Rev. D}\ }\textbf {\bibinfo
  {volume} {100}},\ \bibinfo {pages} {045017} (\bibinfo {year} {2019})},\
  \Eprint {https://arxiv.org/abs/1903.05713} {arXiv:1903.05713 [hep-th]}
  \BibitemShut {NoStop}%
\bibitem [{\citenamefont {Batalin}\ and\ \citenamefont
  {Vilkovisky}(1981)}]{Batalin:1981jr}%
  \BibitemOpen
  \bibfield  {author} {\bibinfo {author} {\bibfnamefont {I.~A.}\ \bibnamefont
  {Batalin}}\ and\ \bibinfo {author} {\bibfnamefont {G.~A.}\ \bibnamefont
  {Vilkovisky}},\ }\bibfield  {title} {\bibinfo {title} {{Gauge algebra and
  quantization}},\ }\bibfield  {booktitle} {\emph {\bibinfo {booktitle}
  {{Second Seminar on Quantum Gravity Moscow, USSR, October 13-15, 1981}}},\
  }\href {https://doi.org/10.1016/0370-2693(81)90205-7} {\bibfield  {journal}
  {\bibinfo  {journal} {Phys. Lett. B}\ }\textbf {\bibinfo {volume} {102}},\
  \bibinfo {pages} {27} (\bibinfo {year} {1981})}\BibitemShut {NoStop}%
\bibitem [{\citenamefont {Elvang}\ and\ \citenamefont
  {Huang}(2013)}]{Elvang:2013cua}%
  \BibitemOpen
  \bibfield  {author} {\bibinfo {author} {\bibfnamefont {H.}~\bibnamefont
  {Elvang}}\ and\ \bibinfo {author} {\bibfnamefont {Y.-t.}\ \bibnamefont
  {Huang}},\ }\bibfield  {title} {\bibinfo {title} {Scattering amplitudes},\
  }\href@noop {} {\  (\bibinfo {year} {2013})},\ \Eprint
  {https://arxiv.org/abs/1308.1697} {arXiv:1308.1697 [hep-th]} \BibitemShut
  {NoStop}%
\bibitem [{\citenamefont {Elvang}\ and\ \citenamefont
  {Huang}(2015)}]{Elvang:2015rqa}%
  \BibitemOpen
  \bibfield  {author} {\bibinfo {author} {\bibfnamefont {H.}~\bibnamefont
  {Elvang}}\ and\ \bibinfo {author} {\bibfnamefont {Y.-t.}\ \bibnamefont
  {Huang}},\ }\href {https://doi.org/10.1017/CBO9781107706620} {\emph {\bibinfo
  {title} {{Scattering amplitudes in gauge theory and gravity}}}}\ (\bibinfo
  {publisher} {Cambridge University Press},\ \bibinfo {year}
  {2015})\BibitemShut {NoStop}%
\bibitem [{\citenamefont {Tyutin}(2002)}]{Tyutin:2000ht}%
  \BibitemOpen
  \bibfield  {author} {\bibinfo {author} {\bibfnamefont {I.}~\bibnamefont
  {Tyutin}},\ }\bibfield  {title} {\bibinfo {title} {{Once again on the
  equivalence theorem}},\ }\href {https://doi.org/10.1134/1.1446571} {\bibfield
   {journal} {\bibinfo  {journal} {Phys. Atom. Nucl.}\ }\textbf {\bibinfo
  {volume} {65}},\ \bibinfo {pages} {194} (\bibinfo {year} {2002})},\ \Eprint
  {https://arxiv.org/abs/hep-th/0001050} {arXiv:hep-th/0001050} \BibitemShut
  {NoStop}%
\bibitem [{\citenamefont {'t~Hooft}\ and\ \citenamefont
  {Veltman}(1974)}]{tHooft:1973wag}%
  \BibitemOpen
  \bibfield  {author} {\bibinfo {author} {\bibfnamefont {G.}~\bibnamefont
  {'t~Hooft}}\ and\ \bibinfo {author} {\bibfnamefont {M.}~\bibnamefont
  {Veltman}},\ }\bibfield  {title} {\bibinfo {title} {{Diagrammar}},\ }\href
  {https://doi.org/10.1007/978-1-4684-2826-1_5} {\bibfield  {journal} {\bibinfo
   {journal} {NATO Sci. Ser. B}\ }\textbf {\bibinfo {volume} {4}},\ \bibinfo
  {pages} {177} (\bibinfo {year} {1974})}\BibitemShut {NoStop}%
\bibitem [{\citenamefont {Leibbrandt}(1975)}]{Leibbrandt:1975dj}%
  \BibitemOpen
  \bibfield  {author} {\bibinfo {author} {\bibfnamefont {G.}~\bibnamefont
  {Leibbrandt}},\ }\bibfield  {title} {\bibinfo {title} {{Introduction to the
  technique of dimensional regularization}},\ }\href
  {https://doi.org/10.1103/RevModPhys.47.849} {\bibfield  {journal} {\bibinfo
  {journal} {Rev. Mod. Phys.}\ }\textbf {\bibinfo {volume} {47}},\ \bibinfo
  {pages} {849} (\bibinfo {year} {1975})}\BibitemShut {NoStop}%
\bibitem [{\citenamefont {Henneaux}\ and\ \citenamefont
  {Teitelboim}(1992)}]{Henneaux:1992}%
  \BibitemOpen
  \bibfield  {author} {\bibinfo {author} {\bibfnamefont {M.}~\bibnamefont
  {Henneaux}}\ and\ \bibinfo {author} {\bibfnamefont {C.}~\bibnamefont
  {Teitelboim}},\ }\href {https://doi.org/10.1515/9780691213866} {\emph
  {\bibinfo {title} {Quantization of gauge systems}}}\ (\bibinfo  {publisher}
  {Princeton University Press},\ \bibinfo {year} {1992})\BibitemShut {NoStop}%
\bibitem [{\citenamefont {Baulieu}(2020)}]{Baulieu:2020obv}%
  \BibitemOpen
  \bibfield  {author} {\bibinfo {author} {\bibfnamefont {L.}~\bibnamefont
  {Baulieu}},\ }\bibfield  {title} {\bibinfo {title} {Unimodular gauge in
  perturbative gravity and supergravity},\ }\href
  {https://doi.org/10.1016/j.physletb.2020.135591} {\bibfield  {journal}
  {\bibinfo  {journal} {Phys. Lett. B}\ }\textbf {\bibinfo {volume} {808}},\
  \bibinfo {pages} {135591} (\bibinfo {year} {2020})},\ \Eprint
  {https://arxiv.org/abs/2004.05950} {arXiv:2004.05950 [hep-th]} \BibitemShut
  {NoStop}%
\end{thebibliography}
    
%

\end{document}